\documentclass[lettersize,journal]{IEEEtran}

\usepackage{amsmath,amsfonts}
\usepackage{caption}
\usepackage{adjustbox}
\usepackage{amsmath}
\usepackage{xspace}
\usepackage{soul}
\usepackage{diagbox}
\usepackage{multirow}
\usepackage{url}
\usepackage{booktabs}
\usepackage{xcolor}
\usepackage{comment}
\usepackage[caption=false,font=normalsize,labelfont=sf,textfont=sf]{subfig}

\usepackage{algorithmicx}
\usepackage[ruled,vlined,linesnumbered]{algorithm2e}
\usepackage{setspace}

\SetKwInput{KwInput}{Input}                
\SetKwInput{KwOutput}{Output}              
\SetKwInput{KwRequire}{Require}            

\graphicspath{{./}{Figure/}}
\newcommand{\parahead}[1]{\vspace*{0.5ex}\noindent %
	{\bfseries #1.}}

\newcommand{\ie}{\emph{i.e.}\xspace}
\newcommand{\eg}{\emph{e.g.}\xspace}

\newcommand{\systemname}{FLAME\xspace}
\newcommand{\slm}{SLM\xspace}
\newcommand{\slms}{SLMs\xspace}

\begin{document}
\title{Taming Asynchronous CPU-GPU Coupling for Frequency-aware Latency Estimation on\\Mobile Edge}
\author{Jiesong Chen, Jun You, Zhidan Liu,~\IEEEmembership{Senior Member,~IEEE,} and Zhenjiang Li,~\IEEEmembership{Member,~IEEE}
\thanks{Jiesong Chen, Jun You, Zhenjiang Li are with the Department of Computer Science, City University of Hong Kong, Hong Kong, China (e-mail: \{jiesochen2-c, junyou3-c\}@my.cityu.edu.hk, \mbox{zhenjiang.li@cityu.edu.hk}).}
\thanks{Zhidan Liu is with INTR Thrust, System Hub, The Hong Kong University of Science and Technology (Guangzhou), Guangzhou, China (\mbox{e-mail:~zhidanliu@hkust-gz.edu.cn}).}
}

\maketitle
\begin{abstract}
Precise estimation of model inference latency is crucial for time-critical mobile edge applications, enabling devices to calculate latency margins against deadlines and trade them for enhanced model performance or resource savings. However, the ubiquity of Dynamic Voltage and Frequency Scaling (DVFS) renders traditional static profiling invalid in real-world deployments, as inference latency fluctuates with varying processor (CPU and GPU) frequencies. While extensive profiling across frequency combinations is theoretically possible, it is prohibitively expensive, particularly for emerging Small Language Models (SLMs), where variable context lengths explode the profiling up to days. We observe that simple analytic scaling fails to predict these fluctuations due to the complex asynchronous coupling between CPU (kernel launching) and GPU (execution). In this paper, we introduce FLAME to accurately estimate inference latency across frequency combinations. It features a novel layer-wise modeling that quantifies the overlapping parallelism and then aggregates dynamic pipeline bubbles caused by asynchronous processor interactions when extending to the full model. This bottom-up approach ensures generalizability across diverse models from DNNs to SLMs, and its precise modeling allows for profiling a sparse subset of samples, cutting DNN profiling from hours to minutes and SLM profiling from days to mere minutes, while maintaining small estimation errors across frequencies. We further showcase FLAME's utility in a deadline-aware DVFS, outperforming the state-of-the-art approach in both power efficiency and latency guarantees.
\end{abstract}

\vspace{-.1in}
\section{Introduction}
\label{sec:1_intro}

Deploying AI models on mobile edge devices, such as the NVIDIA Jetson series and other System-on-Chips (SoCs), requires a delicate balance between performance and resource efficiency~\cite{baccour2022pervasive}. These devices are increasingly crucial to enable innovative, next-generation technologies such as autonomous vehicles~\cite{huang2025modality}, embodied robotics~\cite{ding2025edgemind}, satellite computing~\cite{xing2024deciphering}, unmanned aerial vehicles (UAVs)~\cite{10223251}, etc. 

For time-critical mobile edge applications within these domains, latency awareness is paramount. Specifically, accurately estimating AI model's inference latency \textit{before} execution is a fundamental capability~\cite{nnmeter}. It enables the system to calculate a \textit{residual latency margin} against a reference deadline, serving as the currency to trade excess speed for improved inference quality (via model variant adaptation~\cite{ling2021rt}), reduced power consumption (via frequency scaling~\cite{ztt}), or higher task priority (via preemptive scheduling~\cite{han2024pantheon}), thereby optimizing the overall system or application performance.

The main approach for achieving latency awareness is \textit{profiling-based estimation}~\cite{ling2021rt, han2024pantheon}, where a model is executed offline on the target device to measure its execution latency or other factors (see \S\ref{sec:7_related}) as a prior needed for runtime estimation. While straightforward, existing profiling methods assume that runtime frequencies of CPU and GPU processors match their frequencies used during profiling, typically locking them at maximum levels for consistence~\cite{nnmeter}. However, real-world edge deployments rely on Dynamic Voltage and Frequency Scaling (DVFS) to adjust frequencies dynamically~\cite{choi2019graphics, haj2020sysscale, lin2023workload}, optimizing for thermal and power constraints without sacrificing essential device performance~\cite{ztt}. Thus, static profiling results become invalid once the device adjusts processors' operating frequencies.

A natural question arises: \textit{Can we simply extend profiling to cover all CPU-GPU frequency combinations?} If feasible, it could seamlessly adapt to any DVFS strategy and provide the largest solution space for downstream latency-based optimizations in other resources. While theoretically possible, this \textit{brute-force strategy} is prohibitively expensive. For classic Deep Neural Networks (DNNs) like ResNet, profiling takes tens of minutes to over an hour because low-frequency states greatly extend execution time, and multiple iterations are required to average out jitters~\cite{han2024pantheon}. The challenge intensifies drastically for emerging Small Language Models (SLMs)~\cite{lu2024small}. Unlike DNNs, \slms introduce a third dimension to profiling: context length~\cite{context}, since workload increases with changes in context length, \eg, profiling a Qwen2-7B only up to a 1k context length on a Jetson AGX Orin takes 10+ days (\S\ref{sec:6_eval}). Such overhead renders exhaustive profiling extremely expensive for diverse AI model architectures in practice.

To overcome this issue, an effective solution is to model the latency-frequency relationship analytically. Once this interplay is captured precisely, we can use only a sparse subset of profiling samples to fit latency-frequency modeling parameters and extrapolate to unseen configurations (\eg, frequency combinations, context lengths, etc.), thereby reducing profiling overhead dramatically.

While latency scales inversely with frequency for \textit{isolated processors}~\cite{hennessy2011computer}, this simple linear relationship fails for model inference on mobile edge, because inference on such devices involves complex coupling of the CPU (for preprocessing and kernel launching) and the GPU (for massive parallel computation)~\cite{kim2021minimizing}, which introduces \textit{asynchronous timing dynamics}. The CPU must prepare and feed workloads to the GPU. As the frequencies of the two processors scale, the timing relationship between CPU submission and GPU execution shifts. A frequency configuration that allows perfect pipelining at high frequencies may introduce significant pipeline bubbles (idle waiting times) at lower frequencies. Prior work on single-processor behaviors~\cite{dvfs_ICC, massari2018towards} or coarse modeling/learning~\cite{lyu2024predicting} fails to effectively capture this shifting overlap, leading to large errors when frequencies vary.

In this paper, we introduce \systemname (\underline{F}requency-aware \underline{L}atency \underline{A}nalysis for \underline{M}obile \underline{E}dge) to capture the intricate coupling effects between heterogeneous mobile edge processors. It enables precise latency estimation under arbitrary frequency combinations with very small profiling overhead. To achieve this, we tackle two main challenges:

\textbf{Asynchronous timing dynamics.} The interplay between CPU kernel launching and GPU execution introduces a dynamic ``timing factor'' that varies non-linearly with frequency. This factor represents either the overlapping parallelism or the blocking wait time between the two processors. We propose a dedicated layer-wise formulation that decouples these components, allowing us to build a frequency-dependent modeling to effectively quantify the CPU-GPU overlap under various frequency combinations.

\textbf{Asynchronous pipeline aggregation.} Extending layer-wise estimation to the full AI model is not straightforward, as operating system (OS)-scheduled \textit{asynchrony} allows the CPU to prepare future layers while the GPU executes current ones~\cite{asynchronous}, creating a complex \textit{dependency chain}. Naive summation ignores these pipeline overlaps, leading to cumulative errors. We further introduce a dependency-aware aggregation method that simulates this asynchronous pipeline behavior, ensuring accurate full-model estimation.

By building precise latency-frequency modeling from the bottom up (starting with individual layers and aggregating to the full model), in addition to \textbf{accurate latency estimation}, \systemname delivers other two advantages further:

\textbf{Low profiling overhead.} \systemname only needs to profile unique layers in the model, avoiding redundant measurements of the full model with repeated layers across frequencies. Moreover, the precise latency modeling allows for the use of only sparse subsets of profiling samples, \eg, frequency combinations, context lengths (for \slms), etc. These innovations together dramatically reduce profiling overhead, \eg, shrinking the full profiling time of SLM from days to mere minutes, while maintaining high accuracy.

\textbf{Broad generalizability.} Because \systemname is grounded in layer-wise composition, it is compatible to diverse AI model architectures, ranging from DNNs to \slms. In addition, by explicitly modeling the asynchronous interaction between processors, \systemname supports different CPU-GPU frequency combinations, ensuring seamless integration with different frequency-scaling strategies and also provides ample design space for jointly optimizing latency and other resources (as demonstrated in our design in \S\ref{sec:4_DVFS}).

We implement a prototype of \systemname and conduct extensive evaluations on NVIDIA Jetson AGX Orin and Orin NX platforms across DNNs (ResNet50, VGG16, DenseNet121) and SLMs (GPT2-large, Qwen2-1.5B, Qwen2-7B). The results show that \systemname achieves a latency estimation error of less than 8.14\% across frequencies, while reducing profiling time to 2--6 minutes for DNNs and 2--4 minutes for \slms.

Beyond enabling latency estimation under DVFS, we further design a new DVFS strategy built on \systemname to demonstrate its utility, which retains the lightweight nature of commercial DVFS governors but introduces latency awareness and achieves significantly higher energy efficiency. The DVFS governors on commercial mobile edge optimize power without latency guarantees~\cite{scordino2018energy}. Our enhanced design can dynamically select the most energy-efficient frequency pair that satisfies a specific inference deadline, a critical requirement for video analytics~\cite{bhardwaj2022ekya} and interactive SLM applications~\cite{brysbaert2019many}. Compared to the state-of-the-art learning-based design zTT~\cite{ztt} dedicated for such time-critical applications,\footnote{The more recent GearDVFS~\cite{lin2023workload} outperforms zTT in power efficiency, but latency performance is not guaranteed. Thus, we use zTT for comparison.} our enhanced governor is more lightweight while improving power efficiency by 23.48\% and latency guarantees by 4.35\%. In summary, this paper makes the following contributions:
\begin{itemize}
    \item An analysis of latency estimation pitfalls under frequency scaling, revealing the CPU-GPU asynchrony as the key limiting factor for the estimation.
    \item Novel designs for layer-wise time modeling and pipeline aggregation, enabling precise, low-overhead estimation, alongside a deadline-aware DVFS innovation.
    \item A prototype with comprehensive experiments validating performance improvements over state-of-the-art methods on both classic DNNs and emerging \slms.
\end{itemize}

\section{Background and Motivation}
\label{sec:2_background}

\subsection{Model Inference on Mobile Edge} 
\label{subsec:infer}

\subsubsection{Processor latency and frequency} At the hardware level, the execution latency of a specific processing unit is fundamentally governed by its clock speed~\cite{hennessy2011computer}. Theoretically, for a fixed computational workload, the processing time $t_i$ of a processor $i$ is \textit{inversely proportional} to its operating frequency $f_i$ (\ie, $t_i \propto 1/f_i$)~\cite{he2005power}. However, AI model inference is rarely a single-processor task on mobile edge devices; rather, it relies on heterogeneous CPU-GPU collaboration~\cite{shuvo2022efficient}. 

\subsubsection{CPU-GPU collaborative execution} On mobile edge devices, the CPU functions as the host, responsible for data preprocessing, memory allocation, and launching compute kernels~\cite{yi2023boosting}. The GPU serves as the device/actuator, executing massive parallel computations~\cite{amert2017gpu}.\footnote{The computing paradigm of emerging mobile edge platforms differs from that of conventional mobile devices such as mobile phones. In mobile phones, computation is often CPU-centric with the GPU acting merely as an auxiliary co-processor alongside DSPs and NPUs (see \S\ref{sec:7_related}). Therefore, latency analysis modelings designed for mobile phones~\cite{chen2020deep} are not applicable to the mobile edge context studied in this paper.} Consequently, the \textit{total inference latency is defined as the interval from the moment the CPU initiates the workload to the moment the GPU completes the memory write-back}. 

\begin{figure}[t]
    \begin{center}
        \includegraphics[width=0.99\linewidth]{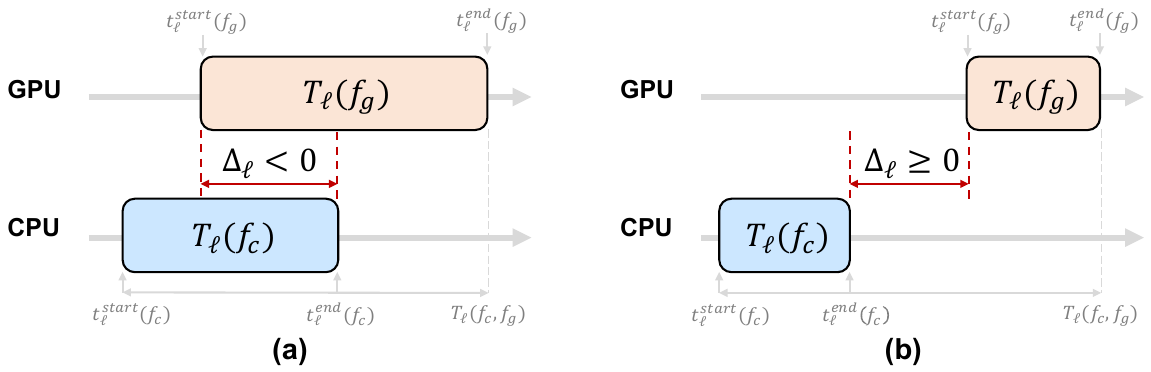}
    \end{center}
    \vspace{-.06in}
    \caption{Dynamic timing factor $\Delta_{\ell}(f_c, f_g)$, denoted as $\Delta_{\ell}$ in the above figure for clear illustration, exists when a mobile edge device processes any model layer $\ell$ due to asynchronous interaction between its CPU and GPU. This asynchrony leads to (a) overlapping and (b) idle waiting between the CPU and GPU execution times.}
    \label{fig:cpugpu}
    \vspace{-.1in}
\end{figure}

\textbf{The dynamic interaction factor.} Due to the asynchronous nature of this collaboration, the latency of any model layer $\ell$ cannot be calculated as a simple summation of CPU and GPU execution time. Instead, it consists of three distinct components: the CPU processing time $T_{\ell}(f_c)$, the GPU processing time $T_{\ell}(f_g)$, and a dynamic interaction term. We formalize this by introducing a dynamic interaction factor $\Delta_{\ell}(f_c, f_g)$:
\begin{equation} \label{eqn:latency}
T_{\ell}(f_c, f_g) = T_{\ell}(f_c) + T_{\ell}(f_g) + \Delta_{\ell}(f_c, f_g),
\end{equation}
which is illustrated in Figure~\ref{fig:cpugpu}. The factor $\Delta_{\ell}(f_c, f_g)$ is highly stochastic. It is influenced by processor frequencies, which dictate kernel launch overheads, memory synchronization barriers, and thread block alignment. We define the implications of this factor as follows:
\begin{itemize}
    \item $\Delta_{\ell}(f_c, f_g) < 0$: The CPU and GPU execution for layer $\ell$ overlaps, effectively hiding latency, as illustrated in Figure~\ref{fig:cpugpu}(a).
    \item $\Delta_{\ell}(f_c, f_g) \ge 0$: The execution involves idle waiting time or stalls between the CPU and GPU, Figure~\ref{fig:cpugpu}(b). This occurs when the CPU frequency is relatively low, which results in a slower CPU preparation of data and launching kernels, causing the GPU to start execution later.
\end{itemize}

\begin{figure}[t]
    \begin{center}
        \includegraphics[width=0.95\linewidth]{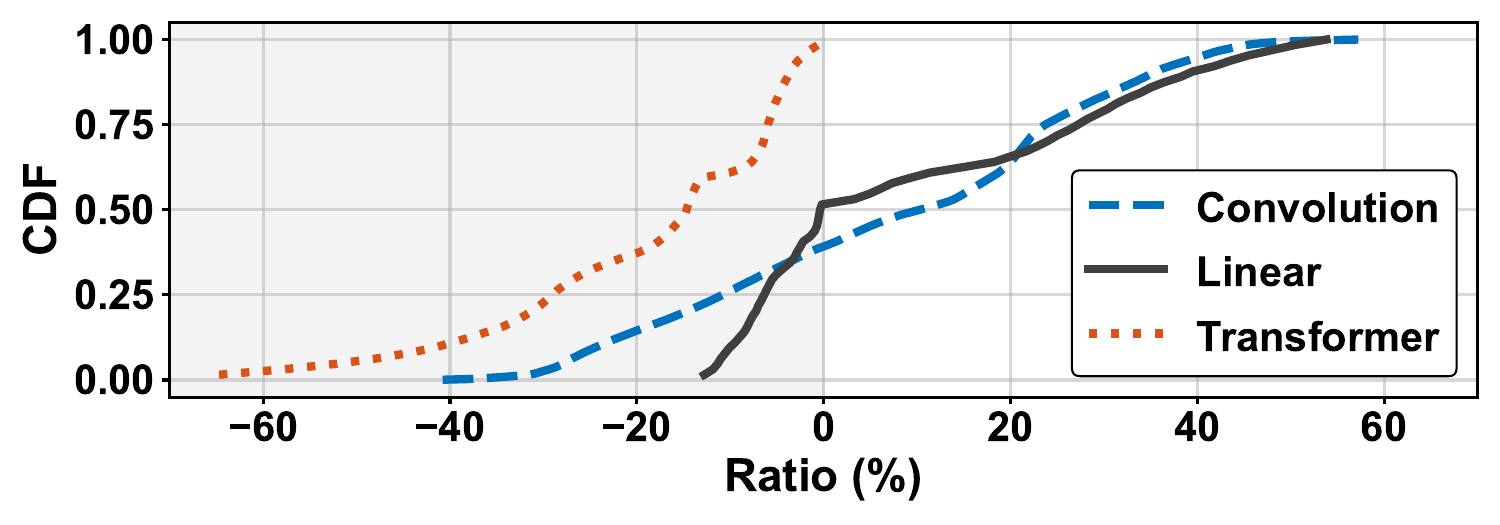}
    \end{center}
    \vspace{-.1in}    
    \caption{CDF of the ratio between the duration of the timing factor $\Delta_{\ell}(f_c, f_g)$ and the execution latency of the corresponding entire layer.}
    \label{fig:2_CDF}
    \vspace{-.1in}
\end{figure}

\textbf{Dynamics of $\Delta_{\ell}(f_c, f_g)$.} To quantify the impact of this factor, we analyze its distribution across various CPU-GPU frequency combinations. Figure~\ref{fig:2_CDF} presents the Cumulative Distribution Function (CDF) of the relative ratio between $\Delta_{\ell}(f_c, f_g)$ and the total layer latency for three typical AI model layers: convolution, linear, and transformer layers.

The results indicate that this factor is highly dynamic and can account for more than 60\% of the total layer latency. For convolution and linear layers, approximately half of the frequency combinations result in execution overlap (\ie, $\Delta_{\ell}(f_c, f_g) < 0$). Conversely, because transformer layers involve a complex sequence of computations, they result in CPU-GPU overlap in nearly all cases. Given the inherently complex nature, it is challenging to analytically derive a precise value for $\Delta_{\ell}(f_c, f_g)$ directly.

\begin{table}[h]
    \centering
    \caption{Overhead of exhaustive profiling for DNN and SLM models on NVIDIA Jetson AGX Orin.}
    \label{tab:profilecost}
    \begin{tabular}{lc|lc}
        \toprule
        \textbf{Model} & \textbf{Time} & \textbf{Model} & \textbf{Time} \\
        \midrule
        ResNet50       & 43 min  & GPT2-large    & 113 h \\
        VGG16          & 54 min  & Qwen2-1.5B    &  151 h \\
        DenseNet121    & 102 min  & Qwen2-7B      & 304 h \\
        \bottomrule
    \end{tabular}
    \vspace{-.1in}
\end{table}

\subsection{Limitations of Existing Methods}

Given the complexity of the interaction described above, we investigate the performance and limitations of existing latency estimation methods.

\subsubsection{Exhaustive profiling} A straightforward solution is to profile the latency across all possible frequency combinations and store the results in a lookup table, ensuring seamless integration with different frequency-scaling strategies and also provides ample design space for jointly optimizing latency and other resources~\cite{hotta2006profile, lai2013latency}. However, the overhead of this approach is prohibitively expensive for practical deployment.

For DNNs, the profiling overhead stems from the combinatorial explosion of the configuration space, because low-frequency states significantly extend execution time, and multiple iterations are required to average out measurement jitter. For instance, the NVIDIA Jetson AGX Orin features 29 CPU and 11 GPU frequency configurations, yielding 319 unique combinations. As shown in Table~\ref{tab:profilecost}, profiling 400 iterations takes nearly an hour in total for standard DNN models, such as 43 minutes for ResNet50, 54 minutes for VGG16 and 102 minutes for DenseNet121 on average.

This overhead is intensified significantly for emerging \slms. Unlike traditional models, \slms introduce an additional dimension to the profiling space: variable context length. Because the computational workload scales with context length, profiling must cover this dimension as well. Table~\ref{tab:profilecost} shows that profiling GPT2-large, Qwen2-1.5B and Qwen2-7B for only up to a 1k context length (5 iterations merely) requires 4.7--12.7 days. This makes exhaustive profiling impractical for real-world scenarios.

\subsubsection{Estimation methods} 
\begin{figure}[t]
    \begin{center}
        \includegraphics[width=0.9\linewidth]{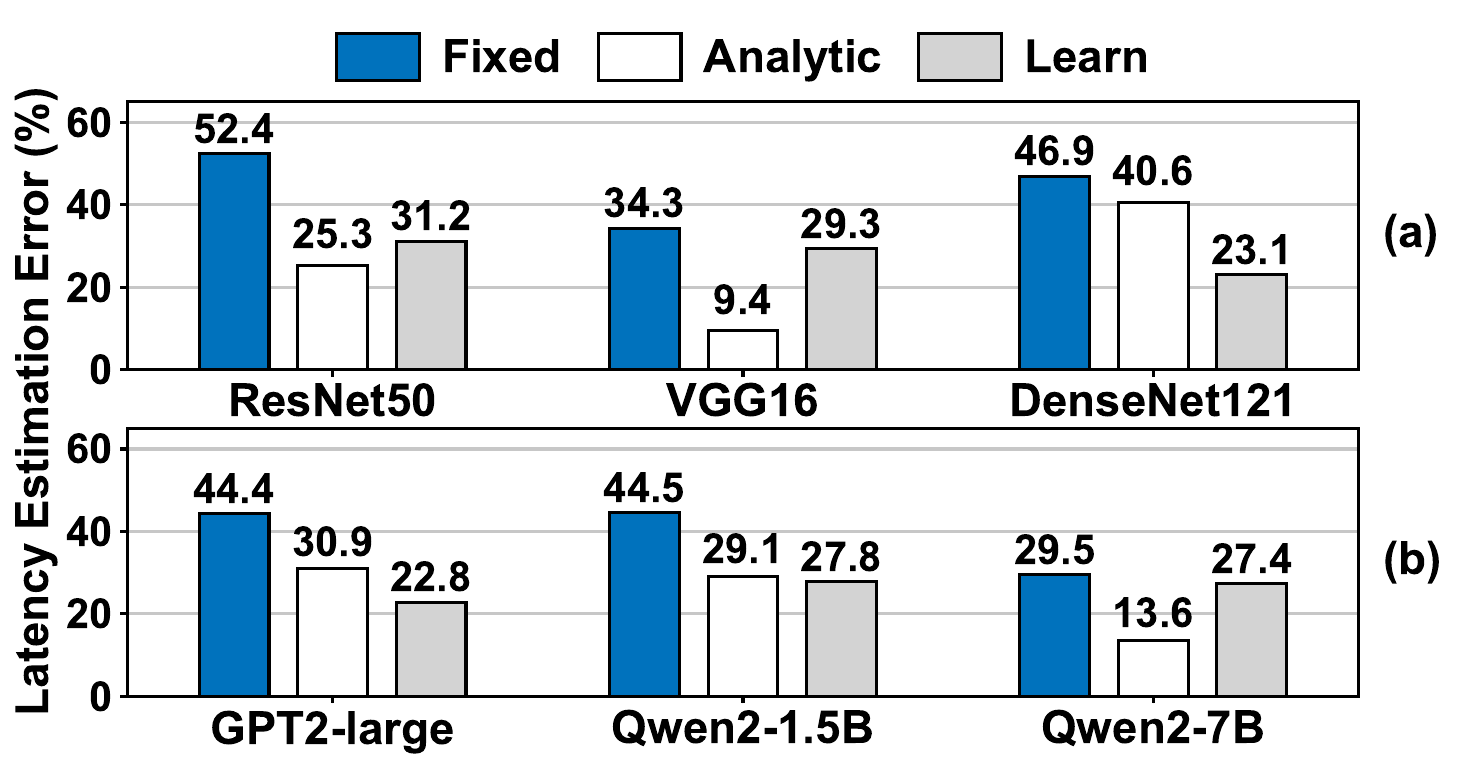}
    \end{center}
    \vspace{-.05in}
    \caption{Latency estimation error of existing methods for (a) DNN models and (b) SLM models.}
    \label{fig:2_motivation_existing}
    \vspace{-.1in}
\end{figure}

Alternative methods attempt to estimate latency without exhaustive profiling, but they struggle to maintain accuracy due to the dynamics of the timing factor $\Delta_{\ell}(f_c, f_g)$. Many existing methods~\cite{nnmeter} assume fixed processor frequencies during both offline profiling and runtime processing. This leads to substantial errors in real-world DVFS-enabled deployments where frequencies fluctuate. As shown in Figure~\ref{fig:2_motivation_existing}, the estimation error of such a method (``Fixed'') averages 44.5\% for DNNs and 39.5\% for SLMs.

Alternatively, analytical methods~\cite{dvfs_ICC} attempt to approximate latency using the standard inverse frequency relationship. However, by failing to account for the interaction factor $\Delta_{\ell}(f_c, f_g)$, the ``Analytic'' method exhibits high error rates, averaging 9.4--40.6\% on DNN models and 13.6--30.9\% on SLMs. Similarly, the learning-based method~\cite{lyu2024predicting} attempts to model the complex relationship in Eq.~(\ref{eqn:latency}) in an end-to-end manner, which also yields substantial errors, \eg, 23.1--31.2\% for DNNs and 22.8--27.8\% for SLMs, primarily because lightweight machine learning models, which are necessitated by system efficiency constraints~\cite{lin2023workload}, struggle to capture the highly dynamic nature of $\Delta_{\ell}(f_c, f_g)$.

\subsection{Design Overview}

\begin{figure}[t]
    \begin{center}
        \includegraphics[width=0.95\linewidth]{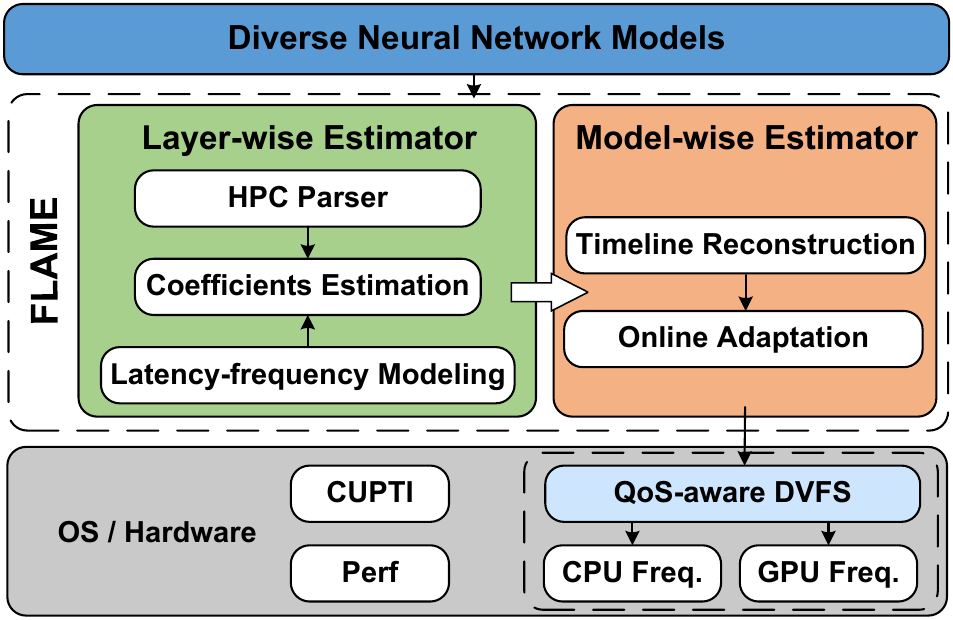}
    \end{center}
    \vspace{-.05in}
    \caption{System architecture of \systemname.}
    \label{fig:3_design_overview}
    \vspace{-.1in}
\end{figure}

To overcome these limitations, we introduce \systemname that delivers accurate latency estimation under dynamic frequency scaling with minimal profiling overhead. Figure~\ref{fig:3_design_overview} depicts the \systemname architecture, which comprises two key components:
\begin{itemize}
    \item \textit{Layer-wise estimator}: It constructs a layer-wise modeling of the target model using only a sparse subset of profiling samples with hardware metadata (\S\ref{sec:3_design_1}).
    \item \textit{Model-wise estimator}: It aggregates the layer-wise latencies to derive the full model latency, explicitly accounting for asynchronous execution overlaps (\S\ref{sec:3_design_2}). 
\end{itemize}

\systemname is a hierarchical approach, which first models per-layer latencies while isolating the timing factor, and then aggregates them into full-model estimates by considering pipeline overlaps and dependencies. This precise modeling requires only sparse profiling samples, making the design efficient and generalizable across DNNs and SLMs.

\section{System Design} \label{sec:3_design}

\subsection{Layer-wise Latency Estimation} \label{sec:3_design_1}

\systemname starts with \textit{layer-wise} latency modeling, making full use of the modular nature of modern AI models, which are typically constructed from a finite set of repeating fundamental blocks (\eg, Convolution, Linear, Transformer). By characterizing these constituent layers individually, we decouple profiling from specific model architectures. This significantly reduces profiling overhead: rather than profiling every new model architecture end to end, we characterize the constituent layers once and aggregate them to derive the inference latency of the entire model.

As formulated in \S\ref{sec:2_background}, the total latency $T_{\ell}$ of a layer $\ell$ is governed by the frequencies of the CPU ($f_c$) and GPU ($f_g$). This relationship is expressed as: $T_{\ell}(f_c, f_g) = T_{\ell}(f_c) + T_{\ell}(f_g) + \Delta_{\ell}(f_c, f_g)$ in Eq.~(\ref{eqn:latency}), which contains the following:
\begin{itemize}
    \item $T_{\ell}(f_c)$ and $T_{\ell}(f_g)$: The \textit{independent} execution latency on the CPU and the GPU, respectively.
    \item $\Delta_{\ell}(f_c, f_g)$: The \textit{dynamic} interaction timing factor between CPU and GPU.
\end{itemize}

\subsubsection{Independent processor time} 
Both CPU and GPU follow a similar principle under frequency scaling --- their processing times are inversely proportional to operating frequency, plus fixed overheads~\cite{dvfs_ICC}. Let $p \in \{c, g\}$ denote CPU or GPU. We model the processor's independent time $T_{\ell}(f_p)$ as:
\begin{equation} \label{eq:processor_time}
T_{\ell}(f_p) = k_{\ell}^p / f_p + b_{\ell}^p,
\end{equation}
where $k_{\ell}^p$ is a workload-dependent coefficient representing the computational complexity of layer $\ell$ on processor $p$. The term $b_{\ell}^p$ captures frequency-independent overheads, like pipeline stalls and cache misses on CPU, or kernel launch latencies and memory transfer delays on GPU~\cite{dvfs_ICC}.

\begin{figure}[b]
    \begin{center}
        \includegraphics[width=0.98\linewidth]{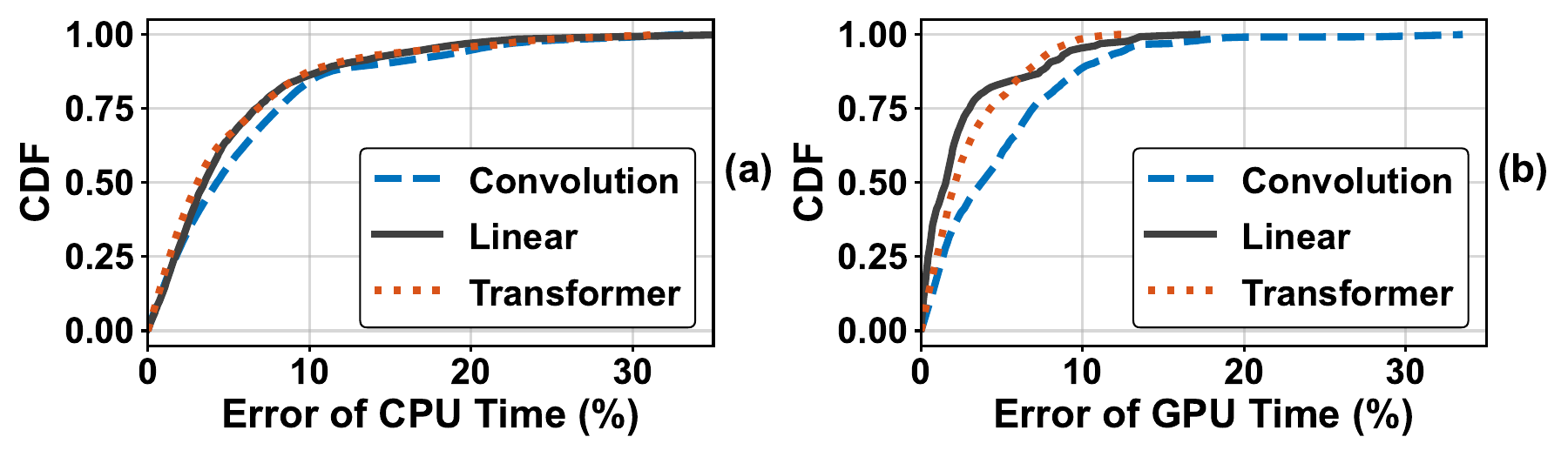}
    \end{center}
    \caption{CDF of estimating independent (a) CPU and (b) GPU times for constituent layers.}
    \label{fig:CPU_GPU}
\end{figure}

To validate the performance, we compare the estimated processing times against ground-truth measurements collected from DNNs and \slms across varying frequencies. As illustrated in Figure~\ref{fig:CPU_GPU}, Eq.~(\ref{eq:processor_time}) demonstrates good accuracy, with over 85\% of the estimation errors falling within 10\% for the CPU and 88\% for the GPU.

\subsubsection{Modeling the dynamic interaction factor} 
The interaction timing factor $\Delta_{\ell}(f_c, f_g)$ captures the temporal coupling between the CPU and GPU during the execution of layer $\ell$. As illustrated in the execution timeline of Figure~\ref{fig:cpugpu}, this factor quantifies the time difference between the GPU's initiation and the CPU's completion:
\begin{equation}
\Delta_{\ell}(f_c, f_g) = t_{\ell}^{start}(f_g) - t_{\ell}^{end}(f_c),
\end{equation}
where $t_{\ell}^{end}(f_c)$ denotes the timestamp when the CPU completes preparatory tasks (\eg, data formatting, kernel launching), and $t_{\ell}^{start}(f_g)$ denotes the timestamp when the GPU begins execution. The factor $\Delta_{\ell}(f_c, f_g)$ serves as an indicator of the hardware parallelism.

\begin{figure}[t]
    \begin{center}
        \includegraphics[width=0.97\linewidth]{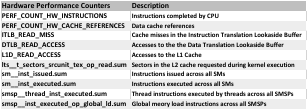}
    \end{center}
    \caption{The selected top-10 HPCs most relevant to the inference latency of the convolution layer.}
    \label{fig:hpc}
    \vspace{-.1in}
\end{figure}

\textbf{Frequency dependency and GPU saturation.} The sign and magnitude of $\Delta_{\ell}(f_c, f_g)$ are strictly governed by the operating frequencies, exhibiting a non-linear phase transition between two distinct states:
\begin{itemize}
    \item \textbf{GPU unsaturated:} When the CPU frequency $f_c$ is relatively low, the system is \textit{CPU-bound}, meaning the CPU cannot issue kernel-launch commands quickly enough to keep the GPU fully occupied. This results in an unsaturated GPU and a positive temporal gap emerges ($\Delta_{\ell}(f_c, f_g) \ge 0$), indicating serial execution pipeline of the workload.
    \item \textbf{GPU saturated:} As the CPU frequency $f_c$ increases, the CPU can prepare data and dispatch kernels more rapidly, allowing the GPU to launch earlier. The system eventually reaches a saturation point where the CPU is no longer the bottleneck. Here, $\Delta_{\ell}(f_c, f_g)$ stabilizes into a negative value ($\Delta_{\ell}(f_c, f_g) < 0$), reflecting maximum pipeline overlap limited only by hardware synchronization overheads.
\end{itemize}

\textbf{Piecewise modeling.} To capture this transition, we employ a piecewise formulation based on a layer-specific saturation frequency threshold, $\hat{f}_{\ell}$. We identify $\hat{f}_{\ell}$ by applying a breakpoint detection algorithm to the profiling data. We then model $\Delta_{\ell}(f_c, f_g)$ as:
\begin{equation} \label{eq:inflect}
\Delta_{\ell}(f_c, f_g) =
\begin{cases}
\frac{k_{\ell, c}^{uns}}{f_c} + \frac{k_{\ell, g}^{uns}}{f_g} + b_{\ell}^{uns}, & \text{if } f_{c} \leq \hat{f}_{\ell}, \\
\frac{k_{\ell, c}^{sat}}{f_c} + \frac{k_{\ell, g}^{sat}}{f_g} + b_{\ell}^{sat}, & \text{if } f_{c} > \hat{f}_{\ell},
\end{cases}
\end{equation}
which differentiates between the two regimes:
\begin{itemize}
    \item In the unsaturated (``uns'') regime ($f_{c} \leq \hat{f}_{\ell}$), the coefficients in Eq.~(\ref{eq:inflect}) capture the dominant impact of CPU processing latency on GPU stall time.
    \item In the the saturated (``sat'') regime ($f_{c} > \hat{f}_{\ell}$), the coefficients in Eq.~(\ref{eq:inflect}) represent the irreducible communication overheads and the maximum degree of CPU-GPU overlap that the hardware can sustain.
\end{itemize}

\subsubsection{Layer-wise estimator} After obtaining all the coefficients $\mathrm{c}_{\ell} = \{k_{\ell}^p, b_{\ell}^p, k_{\ell, c}^{j}, k_{\ell, g}^{j}, b_{\ell}^{j}, \hat{f}_{\ell}\}$, where $j\in \{uns, sat\}$, Eqs.~(\ref{eq:processor_time}) and (\ref{eq:inflect}) form the layer-wise latency estimator for a layer $\ell$, denoted as $\mathrm{est}_{\ell}(f_c, f_g)$. 

\textbf{Determine coefficients.} To derive $\mathrm{c}_{\ell}$, we can measure CPU and GPU processing times, along with overall layer latency, across a sparse set of frequency combinations (\eg, one-sixteenth of all pairs in \S\ref{sec:6_eval}). These measurements enable direct regression fitting of the coefficients $\mathrm{c}_{\ell}$ to instantiate a functional estimator $\mathrm{est}_{\ell}(f_c, f_g)$. 

However, models often contain numerous layers with varying configurations, leading to repeated profiling that increases overhead, even for minor differences. To mitigate this issue, we generalize our design beyond individual layers by profiling representative configurations of each \textit{layer type}, allowing adaptation to hyperparameter variations (\eg, channel depths, kernel sizes, or embedding dimensions). This not only reduces the number of required estimators per model but also minimizes profiling overhead. Furthermore, it supports sharing in a public repository, fostering community contributions of new profiling data for various layer configurations and allowing developers to retrieve them for rapid estimator development or improvement without conducting redundant on-device profiling in the future.

\begin{figure}[t]
    \begin{center}
        \includegraphics[width=0.95\linewidth]{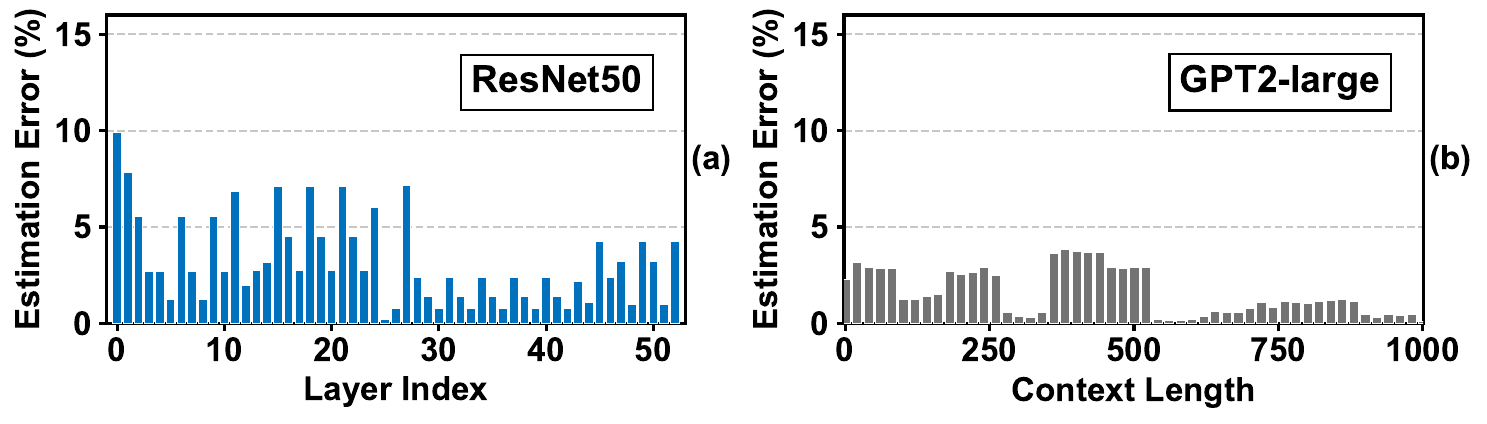}
    \end{center}
    \caption{Estimation error of (a) each layer in ResNet50 and (b) one layer in GPT2-large across context lengths.}
    \label{fig:layer-wiselatency}
    \vspace{-.1in}
\end{figure}

\textbf{Coefficient generalization.} Our key insight for this generalization is that layers of the same type (\eg, Conv2D or Self-Attention) exhibit \textit{structural homogeneity}, \ie, while their hyperparameters differ, their fundamental computational patterns remain similar, differing primarily in workload magnitude. Therefore, if we can identify essential workload features, we can generalize the coefficients $\mathrm{c}_{\ell}$ across different variants of the same layer type.

To quantify these intrinsic features, we use Hardware Performance Counters (HPCs) on mobile edge~\cite{mazzola2022data}. Each HPC counter $m$ exposes an integer $h^m$, which collectively track micro-architectural events like instructions executed, cache misses, and memory accesses~\cite{cuptievent, perf}, providing a high-fidelity fingerprint of the layer workload. To avoid redundancy in the extensive HPC set, we employ Pearson correlation analysis~\cite{liu2015mobile} to identify the most relevant subset for each layer type $\hat{\ell}$, denoted as $\mathrm{HPC}_{\hat{\ell}}$. For efficiency, we retain the top-$n$ ($n=10$ in our implementation), $\mathrm{HPC}_{\hat{\ell}}=\{h_{\hat{\ell}}^m\}_{m=1}^{10}$. Figure~\ref{fig:hpc} details the selected HPCs for the convolution layer.

A practical issue of using HPCs is that they are only accessible \textit{during} or \textit{after} execution, not beforehand for estimating purposes. To enable pre-execution estimation, we build a lightweight XGBoost-based parser~\cite{chen2016xgboost} that maps a layer's static configuration (\eg, input size, embedding dimension, context length) to its expected $\mathrm{HPC}_{\hat{\ell}}$ values. The parser can also be trained during the profiling phase using measured HPCs as ground truth. Therefore, we construct the layer-wise latency estimator for each layer \textbf{type} $\hat{\ell}$ first:
\begin{enumerate}
    \item Input selected configurations of layer type $\hat{\ell}$ (see \S\ref{sec:5_impl}) to the parser to generate $\mathrm{HPC}_{\hat{\ell}} = \{h_{\hat{\ell}}^m\}_{m=1}^{10}$;
    \item Use $\mathrm{HPC}_{\hat{\ell}}$ to regress the coefficients $\mathrm{c}_{\hat{\ell}}$;
    \item Apply $\mathrm{c}_{\hat{\ell}}$ to instantiate the estimator $\mathrm{est}_{\hat{\ell}}(f_c, f_g)$.
\end{enumerate}

Then, we use $\mathrm{est}_{\hat{\ell}}(f_c, f_g)$ as the estimator $\mathrm{est}_{\ell}(f_c, f_g)$ for any individual layer $\ell$, where $\ell$ belongs to the layer type $\hat{\ell}$. In Figure~\ref{fig:layer-wiselatency}(a), we examine the estimation error of each layer in ResNet50, with values ranging 0.19--9.88\% and an average of 3.19\%. In addition, we examine the same Transformer layer in GPT2-large across different context lengths, and the error is also small, such as 0.07--3.87\% in Figure~\ref{fig:layer-wiselatency}(b).

\begin{figure}[t]
    \begin{center}
        \includegraphics[width=0.95\linewidth]{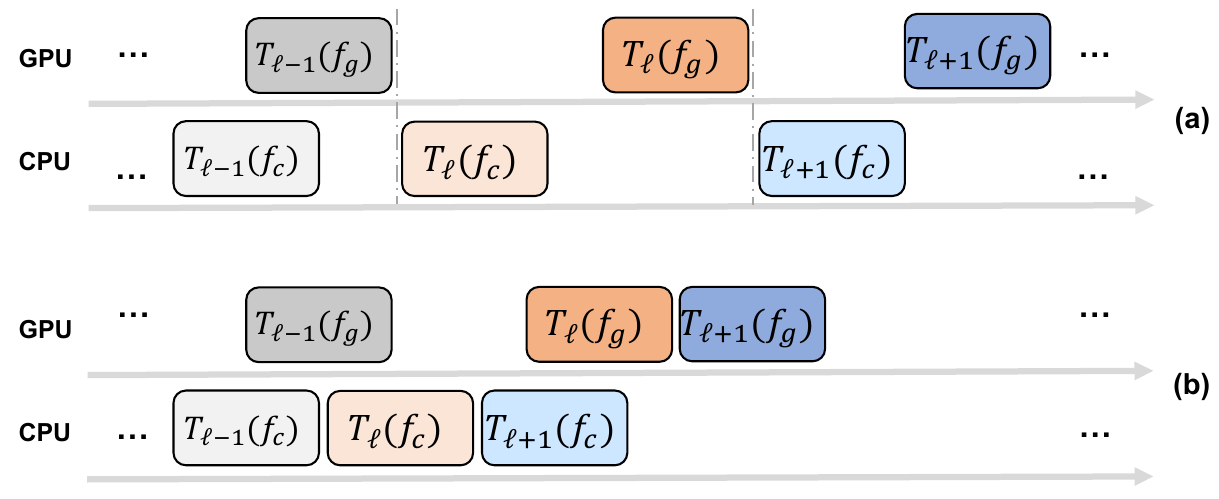}
    \end{center}
    \caption{(a) Ideal serialization and (b) our constructed CPU-GPU processing timeline.}
    \label{fig:3_design_total}
    \vspace{-.1in}
\end{figure}

\subsection{Model-wise Latency Estimation} \label{sec:3_design_2}

Building upon the layer-wise estimator, we extend our methodology to estimate the end-to-end inference latency of complete AI models. This process transits from isolated layer-wise estimation to a comprehensive model-wise analysis, reconstructing the overall execution timeline to account for inter-layer dependencies and the parallelism between CPU and GPU execution.

\subsubsection{Layout-enhanced timeline reconstruction} 
A simplistic method would sum the latencies of all $L$ layers: $T_{model} \approx \sum_{{\ell}=1}^{L} [T_{\ell}(f_c) + T_{\ell}(f_g) + \Delta_{\ell}(f_c,f_g)]$, as shown in Figure~\ref{fig:3_design_total}(a).

However, this summation overlooks the pipelined nature of mobile edge devices, where CPU kernel dispatching and GPU computation often overlap. Operating systems and drivers dynamically schedule tasks to reduce idle bubbles, creating complex timelines that a mere sum fails to capture. To capture these dynamics, we propose a timeline reconstruction method that virtually models the temporal dependencies between processors as follows.

\textbf{CPU timeline.}
In a standard inference pipeline, the mobile-edge CPU functions as the controller, handling data preparation and kernel launching. Since the CPU typically operates asynchronously, dispatching commands to a queue without waiting for the GPU to complete the previous kernel, its timeline is effectively continuous. So, the completion timestamp of the $\ell$-th layer on the CPU is the cumulative sum of its execution time added to the previous timestamp:
\begin{equation} \label{eq:cputimeline}
t_{\ell}^{end}(f_c) = t_{\ell-1}^{end}(f_c) + T_{\ell}(f_c),
\end{equation}
where $t_{\ell}^{end}(f_c)$ is the timestamp the CPU completes layer $\ell$.

\textbf{GPU timeline.} Reconstructing the GPU execution timeline is more intricate than that of the CPU due to two key dependencies: the GPU must wait for kernel dispatch from the CPU and complete the execution of the preceding kernel. To address this, we leverage the more stable CPU timeline along with the dynamic timing factor $\Delta_{\ell}(f_c,f_g)$ to build the GPU timeline. The start time of the GPU for layer $\ell$ is determined as follows based on the value of $\Delta_{\ell}$:
\begin{itemize}
    \item \textit{Case 1: Early GPU Start ($\Delta_{\ell} < 0$).} A negative $\Delta_{\ell}(f_c,f_g)$ indicates the CPU dispatches the kernel early enough for the GPU to begin execution immediately, potentially overlapping with the CPU's tail-end operations:
    \begin{equation}
            t_{\ell}^{start}(f_g) = t_{\ell}^{end}(f_c) - |\Delta_{\ell}(f_c,f_g)|.
    \end{equation}

    \item \textit{Case 2: Delayed GPU Start ($\Delta_{\ell} \geq 0$).} A non-negative $\Delta_{\ell}(f_c,f_g)$ reflects a delay, where the GPU may need to wait for the CPU to complete setup tasks or for the previous GPU kernel to finish. The start time considers both dependencies:
    \begin{equation}
        t_{\ell}^{start}(f_g) = \max \{t_{\ell}^{end}(f_c) + \Delta_{\ell}(f_c,f_g),~t_{\ell-1}^{end}(f_g)\}.
    \end{equation}
\end{itemize}

Finally, the completion time of the GPU for layer $\ell$ is obtained by adding its execution time to the start time: 
\begin{equation}
    t_{\ell}^{end}(f_g) = t_{\ell}^{start}(f_g) + T_{\ell}(f_g).
\end{equation}

Figure~\ref{fig:3_design_total}(b) visualizes this reconstructed timeline. By accounting for these stalls and overlaps, the total inference latency is defined as the span from the CPU's initial start of the first layer to the GPU's completion of the last layer ($L$):
\begin{equation}\label{eq:modeltime}
T(f_c, f_g) = t_L^{end}(f_g) - t_1^{start}(f_c).
\end{equation}

\begin{figure}[t]
    \begin{center}
        \includegraphics[width=0.98\linewidth]{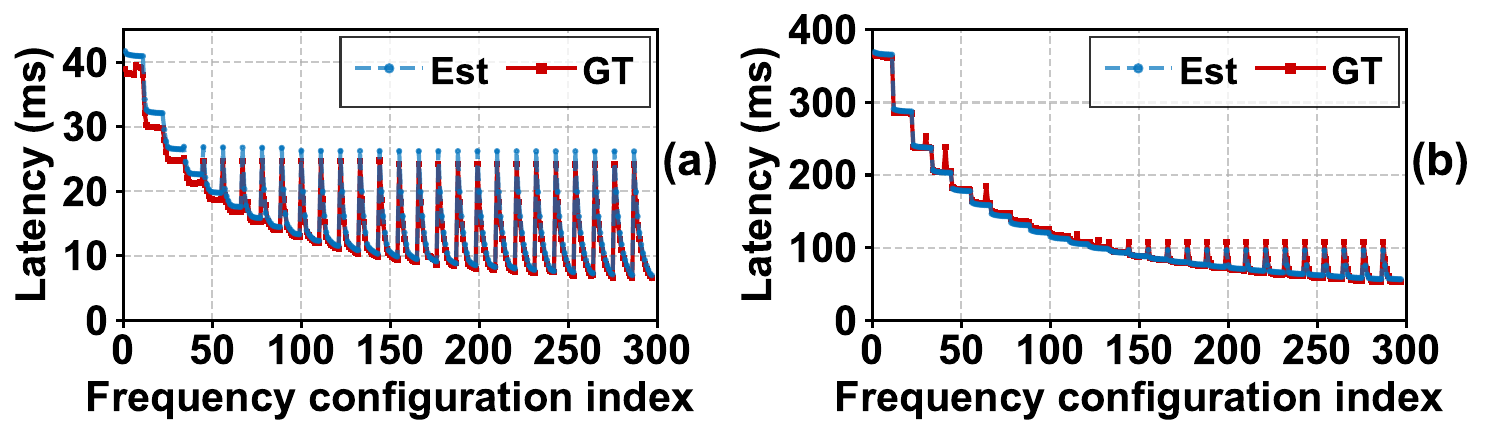}
    \end{center}
    \vspace{.02in}
    \caption{Estimated (Est) model-wise latency of \systemname compared to the ground truth (GT) across all frequency combinations for (a) ResNet50 and (b) GPT2-large.}
    \label{fig:model-wiselatency}
    \vspace{-.1in}
\end{figure}

Figure~\ref{fig:model-wiselatency} validates the performance of our model-wise estimation by sweeping CPU frequencies from lowest to highest, and for each, incrementing GPU frequencies. This allows us to conduct a comprehensive examination. We can see that the estimation error is small, which is 0.01-8.65\% and 0.01-10.9\% for ResNet50 and GPT2-large, respectively.

\subsubsection{Online adaption} While \systemname can generate a model's latency estimates across all frequencies, making them available for integration into other applications, real-world deployments introduce unpredictable disturbances, such as concurrent task loads, resource contention, or OS scheduling jitter, which can skew actual execution times. Without adaptation, these factors lead to accumulating errors in estimations. To maintain robustness, \systemname includes a lightweight online adaptation mechanism that acts as a feedback loop, calibrating estimates against the observed latency values. 

Specifically, we maintain a history of the estimated values $\hat{X}=\{\hat{x}_{i}\}$ along with the corresponding measured latencies $X=\{x_i\}$. We employ a sliding window of size $w+1$ to detect systematic drift by calculating a \textit{local bias} $\sigma_t$ for the current window $t$:
\begin{equation}
\sigma_t = \sum\nolimits_{i=t-w}^{t} (x_i - \hat{x}_i) / (w+1).
\end{equation}
To further smooth transient fluctuations while incorporating historical trends, we apply an exponentially weighted moving average, yielding an adaptive corrector term $\delta_{t}$ by $\delta_t = \alpha \times \delta_t + (1 - \alpha) \times \delta_{t-1}$, where $\alpha \in (0,1]$ (empirically set to 0.6 in our current implementation). During runtime, this term is applied to the model-wise estimation derived in Eq.~(\ref{eq:modeltime}) to produce the final calibrated latency:
\begin{equation}
T(f_c, f_g) = T(f_c, f_g) + \delta_t,
\end{equation}
which is conducted every 10 estimates by non-overlapping historical data. With online adaptation, \systemname sustains estimation accuracy amid dynamic conditions shown in \S\ref{sec:6_eval}.
\section{\systemname-enhanced DVFS Design}
\label{sec:4_DVFS}

\begin{figure}[t]
    \begin{center}
        \includegraphics[height=1.1in]{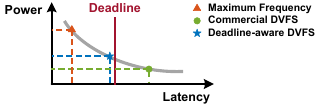}
    \end{center}
    \caption{Deadline-aware DVFS vs.~other strategies.}
    \label{fig:4_DVFS}
\end{figure}

\systemname generalizes latency estimation across CPU and GPU frequencies. This breakthrough can thus transform latency estimation from a \textit{passive metric} into an \textit{active control variable} for time-critical mobile edge computing. 

Fundamentally, \systemname quantifies the \textbf{latency margin} between the estimated inference time and the application deadline, which serves as a resource that can be traded for various system optimizations, such as improved accuracy via model variant adaptation~\cite{ling2021rt}, higher utilization via neural architecture search~\cite{nnmeter}, or enhanced concurrency via preemptive scheduling~\cite{xiang2019pipelined}. Next, we leverage \systemname to develop a lightweight, deadline-aware DVFS that optimizes power efficiency while providing latency guarantees.

\textbf{Motivation.} DVFS is the primary power management mechanism used in mobile devices~\cite{choi2019graphics}. The principle of DVFS is that for any processor, its power consumption $P \propto f \times V^2$~\cite{he2005power}, where $f$ and $V$ are the operating frequency and voltage of the processor, respectively. As $V$ is further proportional to $f$, usually $P \propto f^3$ ~\cite{eyerman2011fine}, DVFS thus adjusts $f$ for power management. Commercial governors (\eg, \textit{schedutil} and \textit{nvhost\_podgov}) typically rely on rule-based heuristics driven by processor utilization. While these strategies are generalizable, they prioritize power reduction and remain agnostic to task-specific latency requirements. Consequently, when handling time-critical workloads, commercial governors often scale frequencies too conservatively, leading to deadline violations~\cite{ztt}.

By integrating \systemname, we enable a ``Just-in-Time'' processing paradigm. Because \systemname estimates the execution time prior to inference, the DVFS governor can proactively select \textit{the most energy-efficient frequency configuration} that satisfies the deadline, as illustrated in Figure~\ref{fig:4_DVFS} (compared to fixed maximum frequencies and commercial strategies). Furthermore, as a general-purpose latency estimator, \systemname enables this DVFS governor to be compatible to diverse model architectures and latency deadlines.

\textbf{Formulation.}
The objective of our design is to minimize the power consumption $P(f_{c}, f_{g})$ during inference, subject to the constraint that the estimated latency $T(f_c, f_g)$ must not exceed the deadline $T_{d}$. This is formulated as:
\begin{equation}
\label{eq:dvfs_opt}
    \min\nolimits_{\{f_{c}, f_{g}\}} P(f_{c}, f_{g}), \quad \textrm{s.t.} \quad T(f_c, f_g) \le T_{d},
\end{equation}
where $f_c$ and $f_g$ represent the CPU and GPU frequencies, respectively. In our current design, the granularity of power governing is to adjust $f_c$ and $f_g$ for the entire model inference for DNNs, and at the per-token level for \slms, to conform to typical application deadline requirements.

\textbf{Design.} Solving Eq.~(\ref{eq:dvfs_opt}) directly is challenging because mobile edge devices typically report only aggregated system power. This makes it infeasible to build a precise, granular modeling to capture the power of individual CPU/GPU and their coupling effects (our tests with end-to-end power estimation yielded high errors). To overcome this, we leverage a trend that lower frequencies yield lower power consumption~\cite{zhang2024improving}. Thus, we can find the \textit{lowest frequency} combination of $f_c$ and $f_g$ to approximate the minimization of $P(f_c, f_g)$. 

However, the search space is non-trivial. For example, the NVIDIA Jetson AGX Orin possesses over 300 valid frequency combinations. Exhaustively estimating latency for every pair to find the global minimum is computationally prohibitive for real-time power governing. To balance optimality with runtime overhead, we propose a decoupled greedy search strategy with linear complexity. Given that the GPU is the primary accelerator for AI workloads on mobile edge, we prioritize identifying the minimum necessary GPU frequency before optimizing the CPU as follows:

First, we pin the CPU to its maximum frequency $f_c^{max}$ to eliminate CPU-side bottlenecks. We then search for the minimum GPU frequency $\hat{f}_g$ that satisfies the deadline:
\begin{eqnarray}
    \hat{f}_g &=& \arg \min\nolimits_{\{f_g\}} \{f_g \mid T(f_c^{max}, f_g) \le T_d\}.
\end{eqnarray}
Second, we fix the GPU frequency to $\hat{f}_g$ and search for the minimum CPU frequency $\hat{f}_c$ that adheres to the deadline:
\begin{eqnarray}
    \hat{f}_c &=& \arg \min\nolimits_{\{f_c\}} \{f_c \mid  T(f_c, \hat{f}_g) \le T_{d}\},
\end{eqnarray}
where the resulting pair of $\hat{f}_c$ and $\hat{f}_g$ is selected for execution. This approach effectively decouples the search space, reducing the complexity from $O(|F_c| \times |F_g|)$ to $O(|F_c| + |F_g|)$, where $|F_c|$ and $|F_g|$ are the number of CPU and GPU frequencies, respectively. As shown in the evaluation, this governor greatly outperforms the state-of-the-art learning-based solution in both energy efficiency and deadline satisfaction.
\section{Implementation}
\label{sec:5_impl}

\parahead{Hardware} 
We implement a prototype of \systemname on the NVIDIA Jetson AGX Orin, which integrates an Arm Cortex‑A78AE-based multi‑core CPU complex together with an NVIDIA Ampere‑architecture GPU, supporting frequency scaling ranges of 0.1--2.2 GHz and 0.3--1.3 GHz, respectively. To evaluate the system's generalization capability, we further deploy and validate \systemname on the NVIDIA Jetson Orin NX, a more resource-constrained edge device.

\textbf{Software.} \systemname is implemented as a lightweight middleware layer positioned between the operating system kernel and the user-space applications. The DNN and \slm inferences are performed using PyTorch~\cite{paszke2019pytorch} and the Transformers library~\cite{wolf2020transformers}. For estimation and profiling, we develop the HPC parser based on XGBoost~\cite{chen2016xgboost} to obtain the selected HPC set for each layer type by inputting the layer's configuration, \eg, the size of the feature map and kernels and the number of input and output channels for convolution; the number of input and output channels for linear; and the context length and embedding dimensions for transformer. The latency estimators are implemented using scikit-learn~\cite{pedregosa2011scikit}. To facilitate efficient HPC access, we develop a custom toolkit that integrates Linux Perf~\cite{perf} for CPU-related counters and the CUDA Profiling Tools Interface~\cite{cuptievent} for GPU-related counters.

To train the HPC parser and all the coefficients $\mathrm{c}_{\ell}$, we profile the constituent layers of the target AI models to collect runtime execution data. We primarily profile convolution and linear layers for DNNs and transformer layers for \slms, as they dominate their computational workload and latency. While other layer types exist, \eg, ReLU activation and batch normalization, their runtime overhead is generally small~\cite{nnmeter}, so we ignore these lightweight layers in our current development. A key feature of \systemname is that it allows the use of a sparse sampling strategy to minimize profiling overhead. We sample the frequency configuration space at an interval of 4 for both CPU and GPU, requiring to profile only $\frac{1}{16}$ ($=6.25\%$) of the total frequency combinations. For \slms, we further use a sampling interval of 90 for context length ($\frac{1}{90}$). During testing, however, we exhaustively evaluate all frequency combinations and context lengths. Power consumption is monitored using the device built-in INA3221 sensor.

\textbf{Models.}
We evaluate \systemname in the next section using a diverse set of representative DNNs (ResNet50, VGG16 and DenseNet121) and \slms (GPT2-large, Qwen2-1.5B and Qwen2-7B). For \slms, we focus on the autoregressive decoding phase because it typically dominates the inference time and therefore demands fine-grained power management. All models are instantiated from the official Torchvision~\cite{marcel2010torchvision} and Transformers~\cite{wolf2020transformers} libraries without architectural modification to ensure reproducibility.
\section{Evaluation}
\label{sec:6_eval}

\begin{figure*}[t]
    \centering
    \begin{minipage}[t]{0.32\textwidth}
        \centering
        \includegraphics[width=\linewidth]{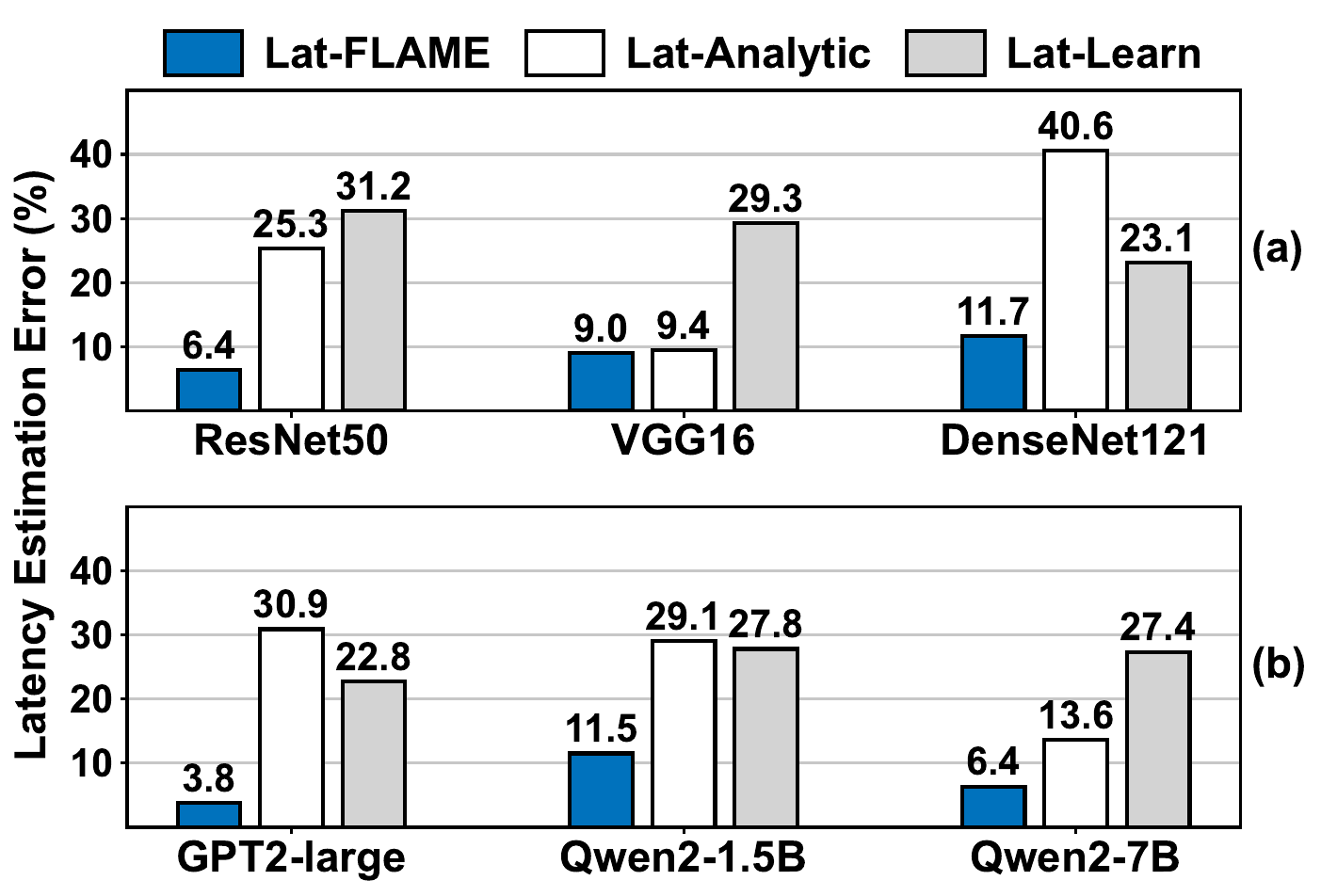}
        \vspace{-.1in}
        \captionof{figure}{Overall inference latency estimation performance on (a) DNN models and (b) SLM models.}
        \label{fig:6_eval_overall_latency_error}
    \end{minipage}
    \hfill
    \begin{minipage}[t]{0.32\textwidth}
        \centering
        \includegraphics[width=\linewidth]{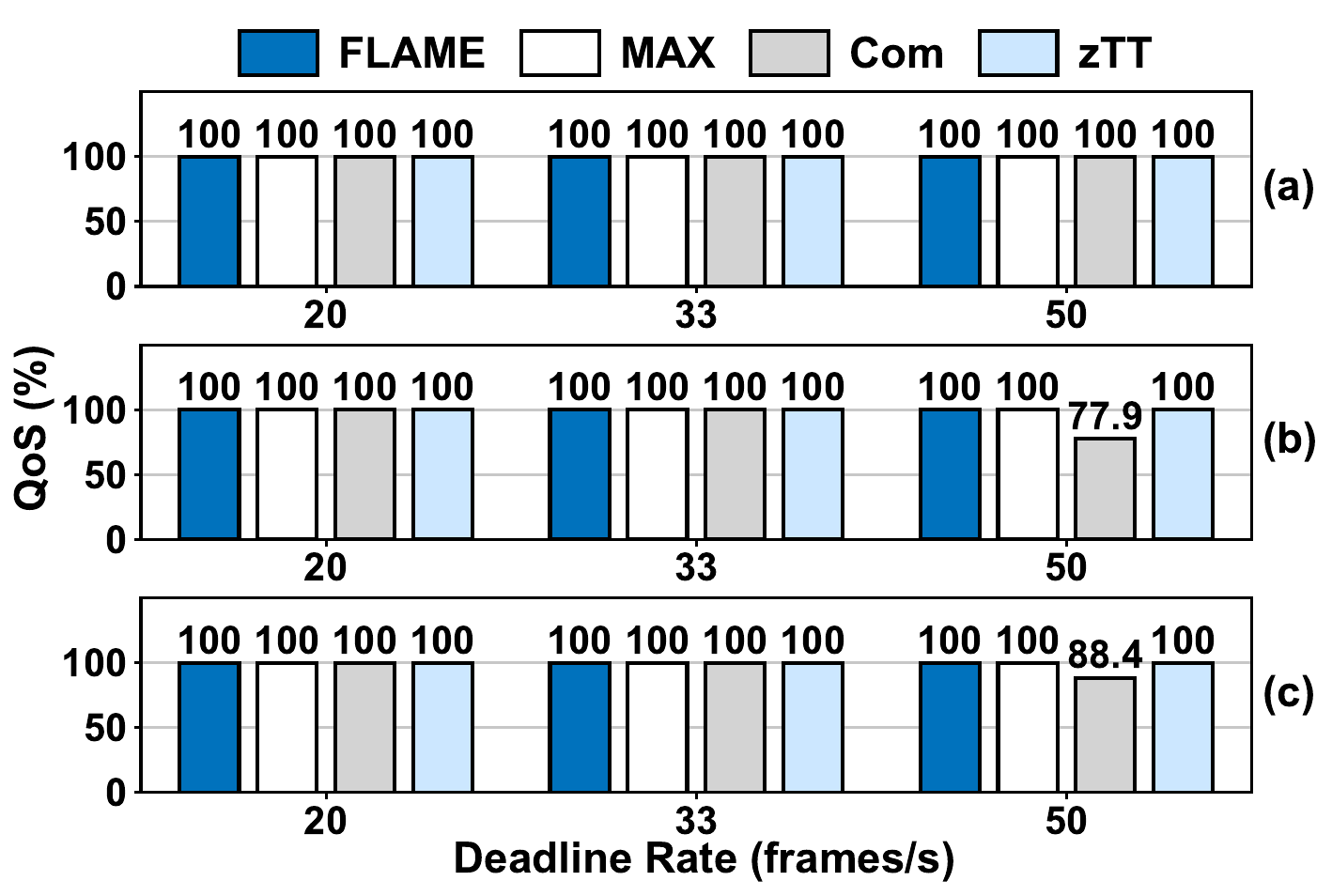}
        \vspace{-.1in}
        \captionof{figure}{QoS of governors at different deadline rates on (a) ResNet50 (b) VGG16 and (c) DenseNet121.}
        \label{fig:6_eval_overall_QoS_DNN}
    \end{minipage}
    \hfill
    \begin{minipage}[t]{0.32\textwidth}
        \centering
        \includegraphics[width=\linewidth]{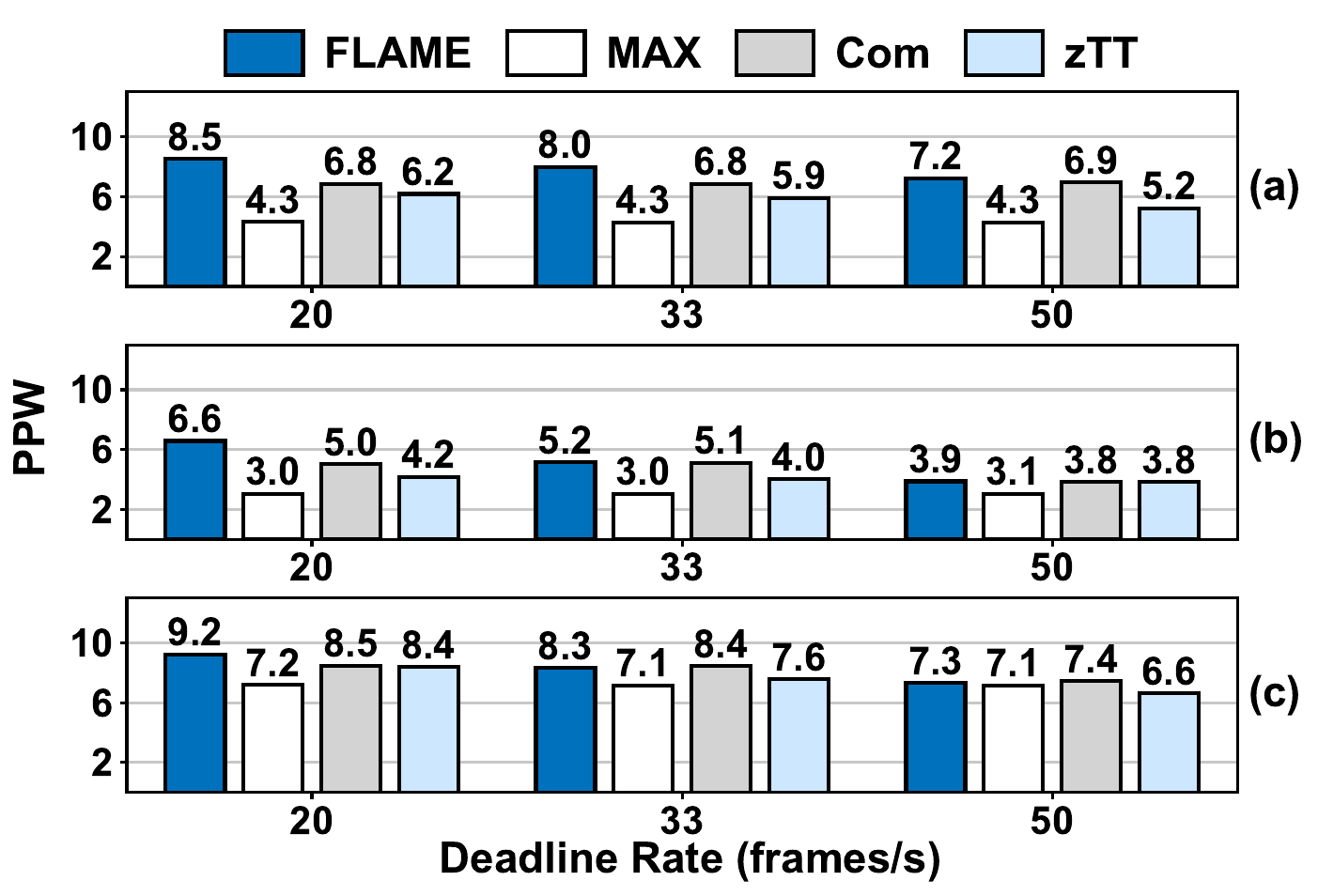}
        \vspace{-.1in}
        \captionof{figure}{PPW of governors at different deadline rates on (a) ResNet50, (b) VGG16 and (c) DenseNet121.}
        \label{fig:6_eval_overall_PPW_DNN}
    \end{minipage}
    \vspace{-.1in}
\end{figure*}

\subsection{Experimental Setups} 

\subsubsection{Baselines} We evaluate the performance and effectiveness of \systemname from two perspectives: \textit{latency estimation performance} and \textit{end-to-end DVFS performance}.

\textbf{1) Latency estimation performance.} We compare \systemname (denoted as \textbf{Lat-\systemname}) against the following methods:

\begin{itemize}
    \item \textbf{Deep Learning (Lat-Learn)}: A data-driven, end-to-end latency estimator based on MLPs~\cite{lyu2024predicting}.

    \item \textbf{Analytical (Lat-Analytic):} A parametric curve-fitting approach that models the inverse latency-frequency relationship as $T = a \cdot f_{g}^{-b} + c$, where $a$ to $c$ are learnable coefficients~\cite{dvfs_ICC}.

\end{itemize}

\textbf{2) DVFS performance.} We further compare the \systemname-enhanced DVFS governor with the following governors:

\begin{itemize}
    \item \textbf{Maximum Frequency (DVFS-MAX):} A static policy that fixes both CPU and GPU to their maximum frequencies to minimize latency, serving as the upper bound for power consumption and the lower bound for latency.

    \item \textbf{Commercial (DVFS-Com):} A commercial DVFS governor designed for mobile edge devices, which manages static frequency thresholds for the GPU (via \texttt{nvhost\_podgov}) and the CPU (via \texttt{schedutil})~\cite{schedutil}.

    \item \textbf{zTT (DVFS-zTT):} A state-of-the-art reinforcement learning-based baseline~\cite{ztt} that adapts frequency scaling based on QoS (\ie, latency in our evaluation).

\end{itemize}

\subsubsection{Evaluation Metrics} We compare different methods using the following performance metrics:

\textbf{1) Latency estimation accuracy metric.} The accuracy is measured using the \textit{Mean Absolute Percentage Error (MAPE)} across all frequency configurations: $\textit{MAPE} = \frac{1}{n} \sum_{i=1}^{n} \left| \frac{y_i - \hat{y}_i}{y_i} \right| \times 100\%$, where $n$ denotes the total number of frequency configurations, $y_i$ and $\hat{y}_i$ represent the actual and estimated latencies, respectively.

\textbf{2) DVFS performance metrics.} The following two metrics are used to evaluate different DVFS governors:

\begin{itemize}
    \item \textbf{Quality of Service (QoS)} measures the success rate of a method to meet predefined deadline-related requirements (\eg, frames or tokens per second). It is quantified as the ratio of \textit{actual performance} ($r_a$) to \textit{desired or required performance} ($r_d$) for each method, capped at 100\%: $\textit{QoS} = \min\left(r_{a}/r_{d}, 1\right) \times 100\%$. A higher QoS value indicates better latency guarantees.

    \item \textbf{Performance per Watt (PPW)} evaluates a system's energy efficiency by measuring the useful work (QoS) delivered per unit of power consumed: $\textit{PPW} = \textit{QoS}/P_{avg}$, where $P_{avg}$ is the average power consumption. A higher PPW value indicates better power efficiency. 
\end{itemize}

\begin{figure*}[t]
    \centering
    \begin{minipage}[t]{0.32\textwidth}
        \centering
        \includegraphics[width=\linewidth]{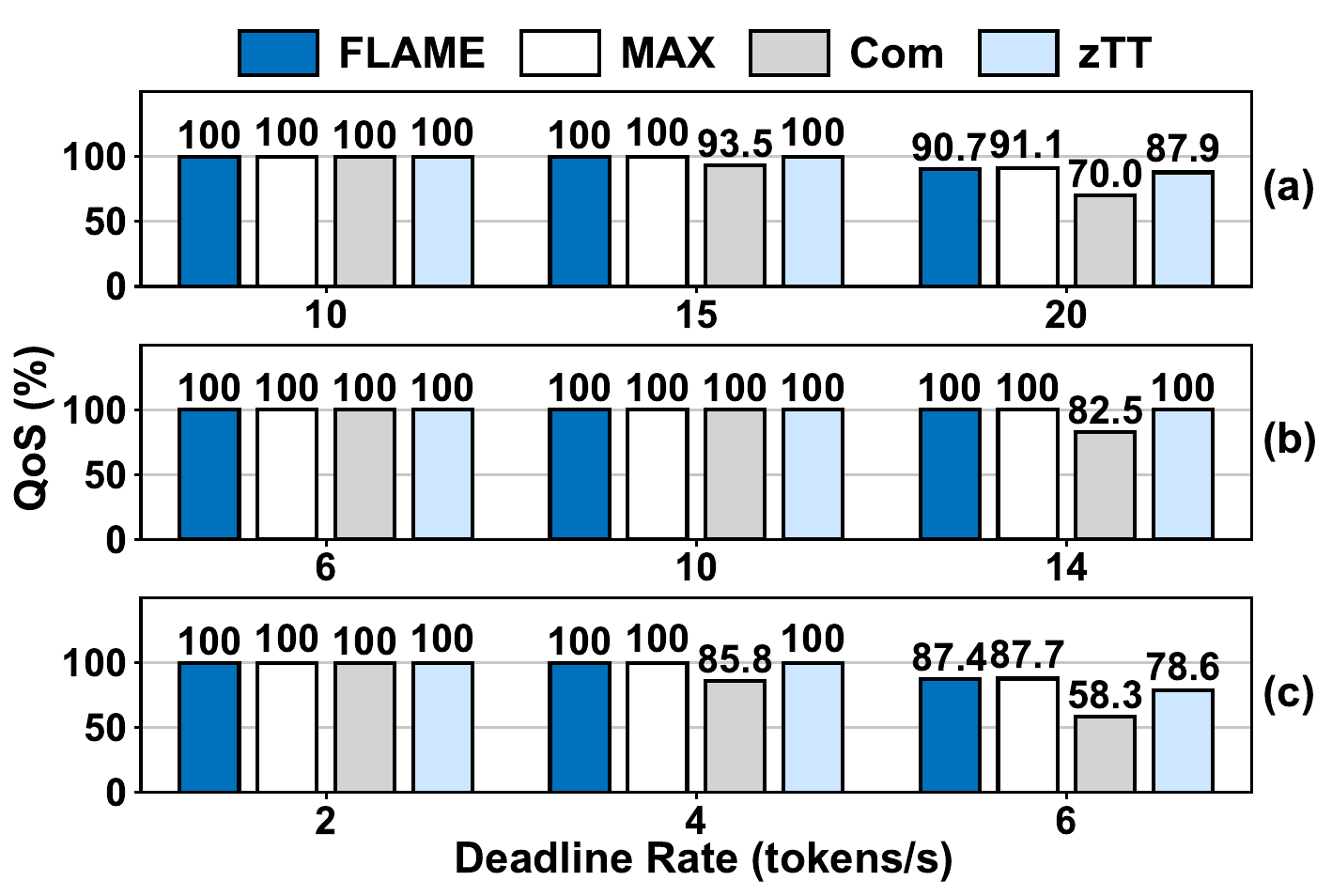}
        \vspace{-.1in}
        \captionof{figure}{QoS of governors at different deadline rates on (a) GPT2-large, (b) Qwen2-1.5B and (c) Qwen2-7B.}
        \label{fig:6_eval_overall_QoS_LLM}
    \end{minipage}
    \hfill
    \begin{minipage}[t]{0.32\textwidth}
        \centering
        \includegraphics[width=\linewidth]{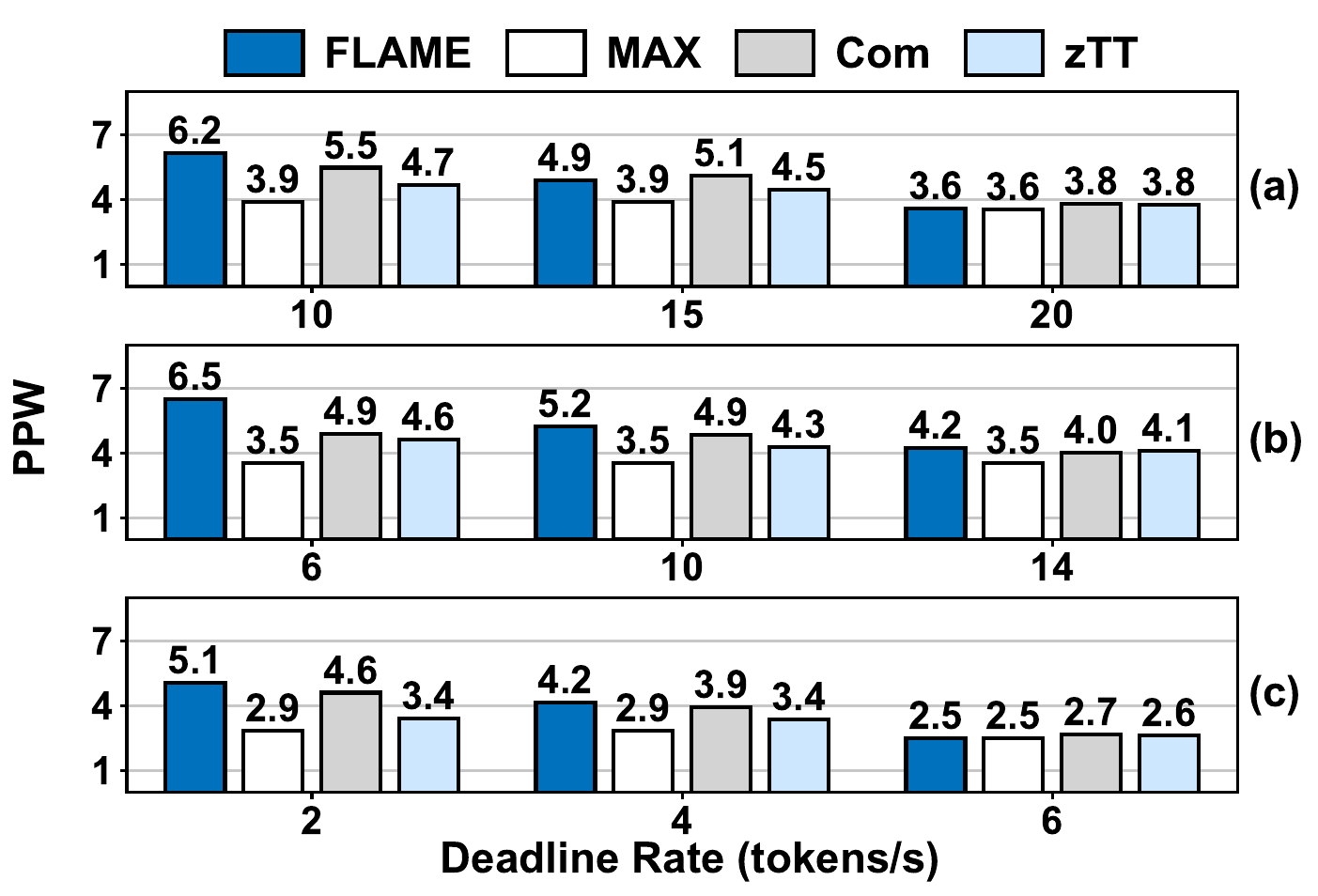}
        \vspace{-.1in}
        \captionof{figure}{PPW of governors at different deadline rates on (a) GPT2-large, (b) Qwen2-1.5B and (c) Qwen2-7B.}
        \label{fig:6_eval_overall_PPW_LLM}
    \end{minipage}
    \hfill
    \begin{minipage}[t]{0.32\textwidth}
        \centering
        \includegraphics[width=\linewidth]{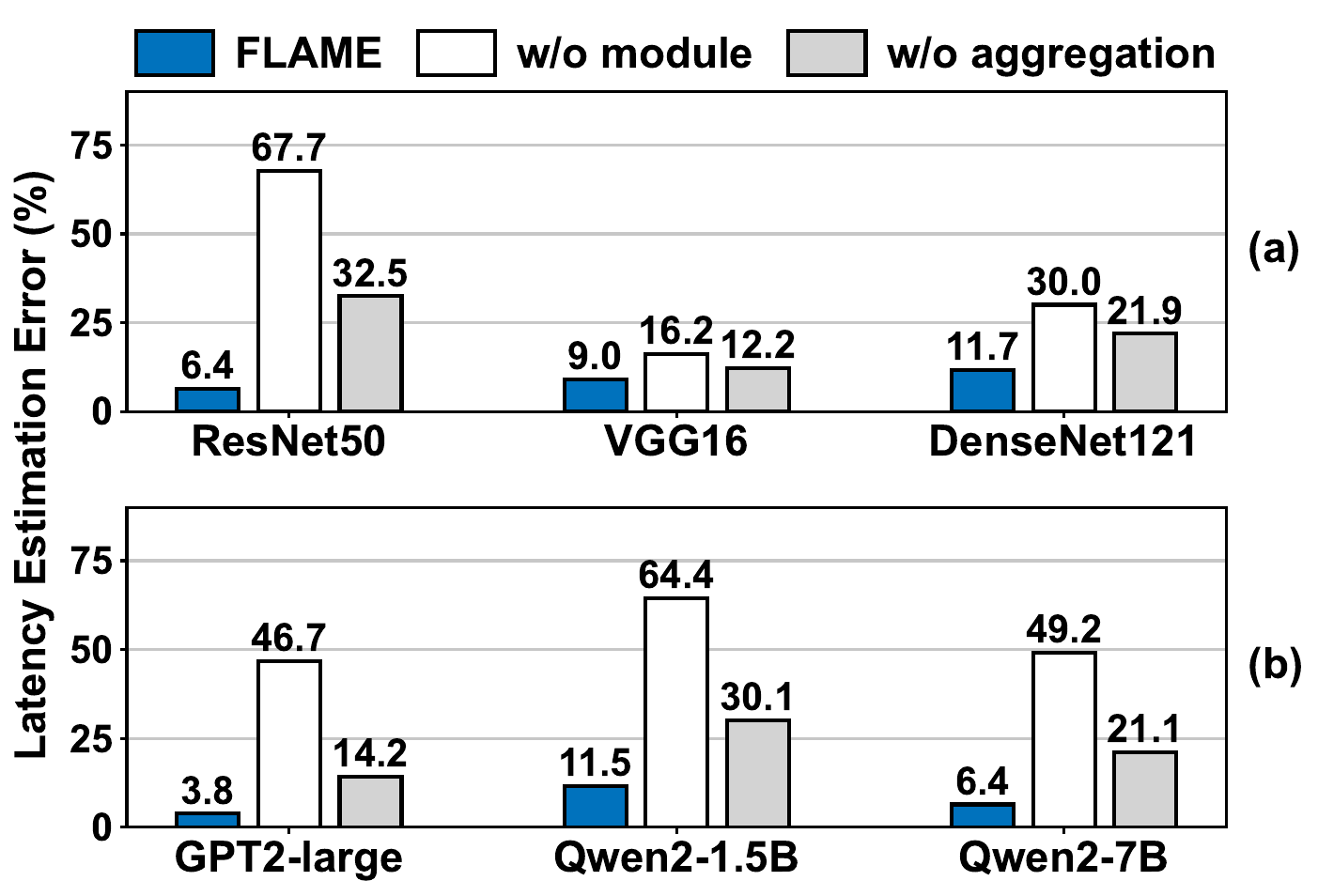}
        \vspace{-.1in}
        \captionof{figure}{Ablation study of different FLAME variants on (a) DNN models and (b) SLM models.}
        \label{fig:6_eval_ablation}
    \end{minipage}
    \vspace{-.1in}
\end{figure*}

\subsection{Overall Performance}
\label{s:eval:overall}

\textbf{Latency estimation performance.} We first evaluate the estimation accuracy of \systemname compared to other two baseline methods across all the DNN and SLM models, covering all frequency combinations and context lengths (up to 1024 for SLMs). Figure~\ref{fig:6_eval_overall_latency_error}(a) and (b) summarize the results for DNN and SLM models, respectively. Both baselines exhibit large estimation errors due to their design limitations. Lat-Analytic, which assumes a simplified linear relationship between frequency and latency, yields an average error of 24.82\%. It performs adequately only on VGG16 (in Figure~\ref{fig:6_eval_overall_latency_error}(a)), where the workload is heavily compute-bound on the GPU, masking the non-linear CPU-GPU interactions that Lat-Analytic ignores. Lat-Learn attempts to capture these non-linearities via regression but fails to generalize effectively to unseen frequency configurations, resulting in a high average error of 26.93\%. In contrast, by explicitly modeling the dynamic coupling between CPU and GPU and accounting for frequency-dependent effects, Lat-\systemname reduces the average estimation error to 8.14\%, outperforming two baselines by 67.23\% and 69.80\%, respectively.

\textbf{DVFS performance.} Next, we evaluate the effectiveness of our enhanced DVFS design compared to other governors.

\textit{DNN workloads.} We first evaluate the governors by executing DNN models under varying Frames Per Second (FPS) deadline constraints from 20 to 50. Figure~\ref{fig:6_eval_overall_QoS_DNN} shows that most methods achieve near 100\% latency QoS rates due to the relatively low computational intensity of these DNN models on AGX Orin. The notable exception is the commercial strategy DVFS-Com, which achieves only 77.93\% QoS for VGG16 and 88.43\% for DenseNet121 at a 50 FPS target due to its inherent latency-agnostic limitation. Regarding power efficiency, we further report their PPW in Figure~\ref{fig:6_eval_overall_PPW_DNN}. As expected, DVFS-MAX (fixed maximum frequency) yields the lowest PPW. While DVFS-Com improves upon this, it does so by occasionally violating deadlines. The state-of-the-art, zTT, generally meets deadlines but lacks precise latency modeling, leading to suboptimal frequency selection. \systemname, conversely, identifies the energy-optimal frequency combination that precisely satisfies the deadline and outperforms zTT by an average of 23.69\% in PPW. 

\textit{SLM workloads.} We further evaluate these methods on SLM workloads, where the metric is token generation rate (tokens per second). Due to the significant variance in computational workloads among diverse SLMs, we assign model-specific deadline rates. Figures~\ref{fig:6_eval_overall_QoS_LLM} and \ref{fig:6_eval_overall_PPW_LLM} present the QoS and PPW results, respectively. The trends observed in DNNs persist here. DVFS-MAX remains the least power efficient, while DVFS-Com frequently fails to satisfy the high computational demands of \slms, resulting in QoS violations. Although zTT satisfies most QoS targets, it lacks the granularity to minimize power usage effectively. In contrast, \systemname operates most efficiently, outperforming DVFS-MAX, DVFS-Com and zTT by 40.39\%, 7.61\% 23.26\% on average in PPW, respectively.

\textbf{Ablation study.} To assess the effectiveness of key technical components, we develop two ablated variants of \systemname and compare them with the full version of \systemname:

\begin{itemize}
    \item \textbf{``w/o module''}: This variant ignores the the dynamic interaction factor $\Delta_{\ell}$ and thereby removes the layer-wise estimator. The final $T$ is simply the sum of CPU and GPU processing time of all constituent layers.
    
    \item \textbf{``w/o aggregation''}: This variant adopts the layer-wise estimator to estimate per-layer latency, but aggregates these predictions as the final $T$ by simple summation.
\end{itemize}

\begin{figure*}[ht]
    \centering
    \begin{minipage}[t]{0.32\textwidth}
        \centering
        \includegraphics[width=\linewidth]{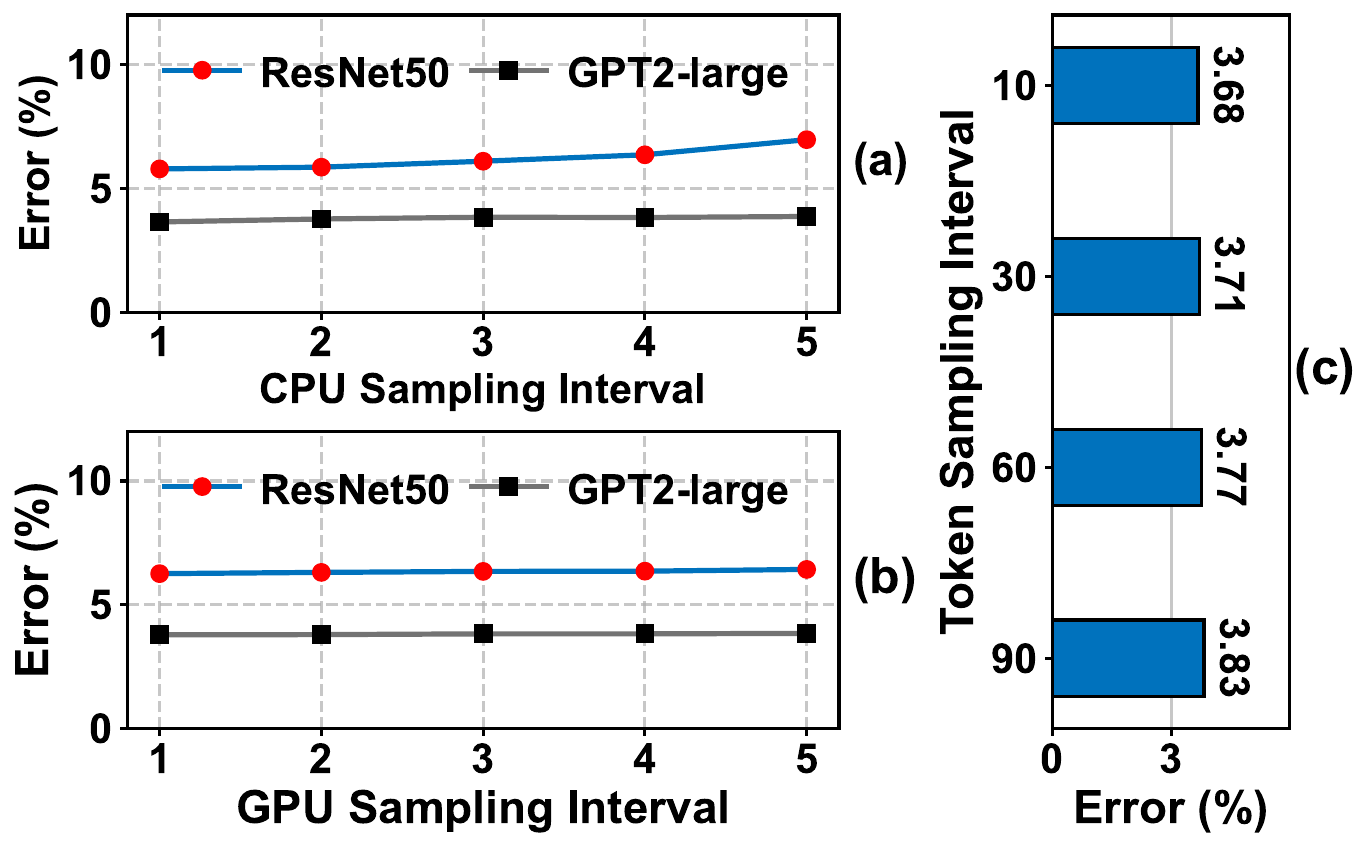}
        \vspace{-.1in}
        \captionof{figure}{Impact of sampling interval of (a) CPU and (b) GPU frequency, and (c) context length of GPT2.}
        \label{fig:6_eval_micro_diff_sample_intervals}
    \end{minipage}
    \hfill
    \begin{minipage}[t]{0.32\textwidth}
        \centering
        \includegraphics[width=\linewidth]{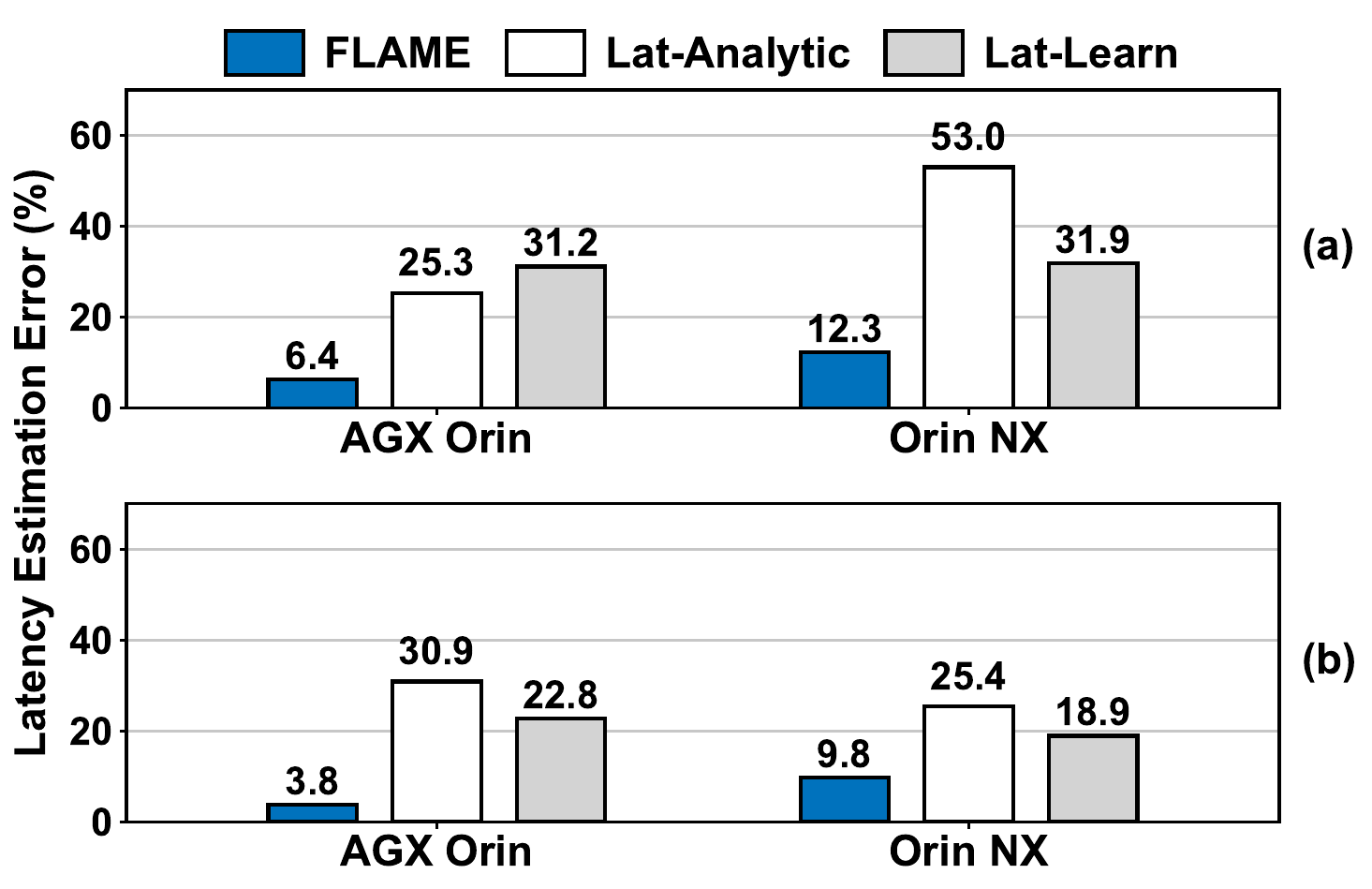}
        \vspace{-.1in}
        \captionof{figure}{Latency estimation error of (a) ResNet50 and (b) GPT2-large on AGX Orin and Orin NX.}
        \label{fig:6_eval_micro_diff_estimator_dual_device}
    \end{minipage}
    \hfill
    \begin{minipage}[t]{0.32\textwidth}
        \centering
        \includegraphics[width=\linewidth]{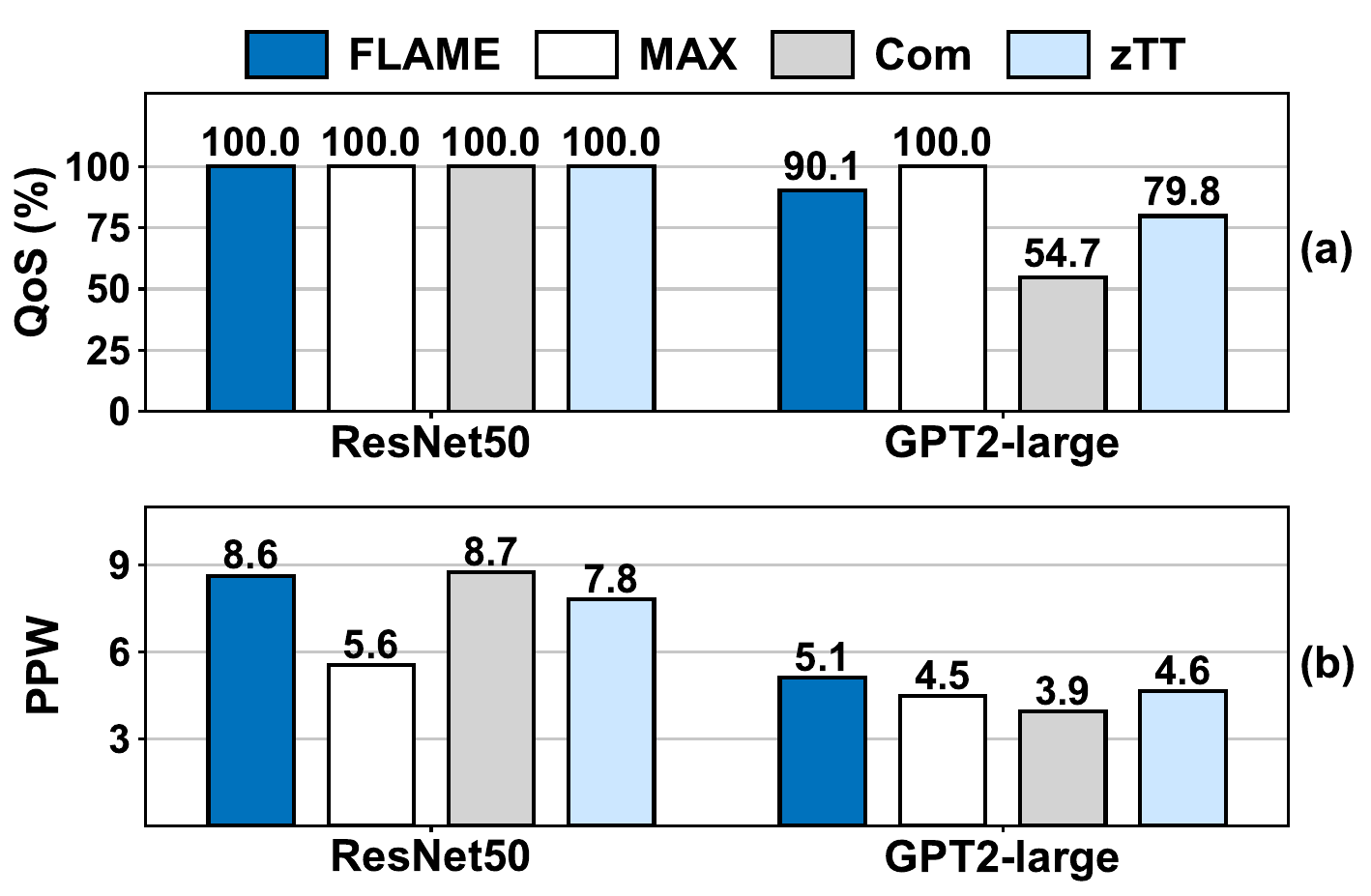}
        \vspace{-.1in}
        \captionof{figure}{Performance comparisons of different DVFS governors in the metric of (a) QoS and (b) PPW.}
        \label{fig:6_eval_micro_diff_devices_NX_only}
    \end{minipage}
    \vspace{-.1in}
\end{figure*}


Figure~\ref{fig:6_eval_ablation}(a) and (b) present the latency estimation errors for both variants on DNN and SLM models, respectively. The variant ``w/o module'' exhibits prohibitively high errors on both kinds of models. It confirms that simple frequency-scaling laws are insufficient for modern AI models, particularly for SLMs that involve more complicated CPU-GPU interactions. The variant ``w/o aggregation'', while utilizing layer-wise latency modeling yet without handling inter-layer dependencies and the CPU-GPU parallelism, still shows substantial errors, indicating that knowing individual layer latency is necessary but not sufficient. In contrast, by integrating both precise layer-wise latency estimation and model-wide timeline aggregation, the complete \systemname design achieves significantly higher accuracy.
Compared to the two variants, the complete design can reduce the estimation error by 65.33\% and 51.04\%, respectively, on DNN models, and by 86.99\% and 68.23\%, respectively, on SLM models.

\subsection{Micro-benchmarks}
In this subsection, we conduct micro-benchmark experiments to evaluate \systemname under different settings.

\textbf{Sampling interval.} In this experiment, we investigate how the sampling interval used to profile the frequencies of CPU and GPU and the context length of \slms during the profiling affect the performance. If the latency modeling is accurate, it allows to use fewer profiling samples to train the estimator, which reduces the profiling overhead. Due to the page limitation, we show the results on ResNet50 and GPT2-large (we observe similar results in other models). Figure~\ref{fig:6_eval_micro_diff_sample_intervals} shows that widening the CPU sampling interval from 1 (full profiling) to 4 for \systemname increases errors only modestly, from 5.79\% to 6.97\% for ResNet50 and 3.65\% to 3.87\% for GPT2-large in Figure~\ref{fig:6_eval_micro_diff_sample_intervals}(a). GPU sampling shows even less impact as illustrated in Figure~\ref{fig:6_eval_micro_diff_sample_intervals}(b). 

Based on this robustness, we adopt a uniform interval of 4 for both CPU and GPU by default during the profiling, shrinking the profiling space by $\frac{1}{16}$ ($=\frac{1}{4 \times 4}$) without sacrificing accuracy too much. For \slms, larger token sampling intervals also maintain the performance, as shown in Figure~\ref{fig:6_eval_micro_diff_sample_intervals}(c), and we select the interval of 90 by default (\ie, $\frac{1}{90}$) when profiling context length of \slms in the evaluation.

\textbf{Different devices.} In this experiment, we deploy \systemname on the more resource-constrained Jetson Orin NX to evaluate its generality and compare its latency estimation accuracy against the baselines for ResNet50 and GPT2-large in Figure~\ref{fig:6_eval_micro_diff_estimator_dual_device}, in which we also plot their performance on the high-end AGX Orin as reference. \systemname consistently delivers better estimations on both platforms, with a minor error uptick on the Orin NX due to its constrained resources amplifying OS jitter and background interference. Nonetheless, \systemname still greatly outperforms all baselines on both devices.

We further evaluate different DVFS governors on the Jetson Orin NX platform using ResNet50 and the more computationally demanding GPT2-large. Figure~\ref{fig:6_eval_micro_diff_devices_NX_only}(a) and (b) compare their QoS and PPW performance, respectively. For ResNet50, all governors satisfy QoS target by 100\%, albeit with notable differences in energy efficiency. Among them, \systemname and DVFS-Com achieve comparable energy efficiency. However, their performance diverges significantly on the larger GPT2-large model. While DVFS-MAX continues to meet QoS requirements for the \slm, it does so with substantially higher power consumption. In contrast, \systemname maintains a better balance, delivering high QoS with improved PPW. These results validate the general applicability of \systemname on resource-constrained mobile edge devices.

\begin{figure}[t]
    \begin{center}
        \includegraphics[width=0.97\linewidth]{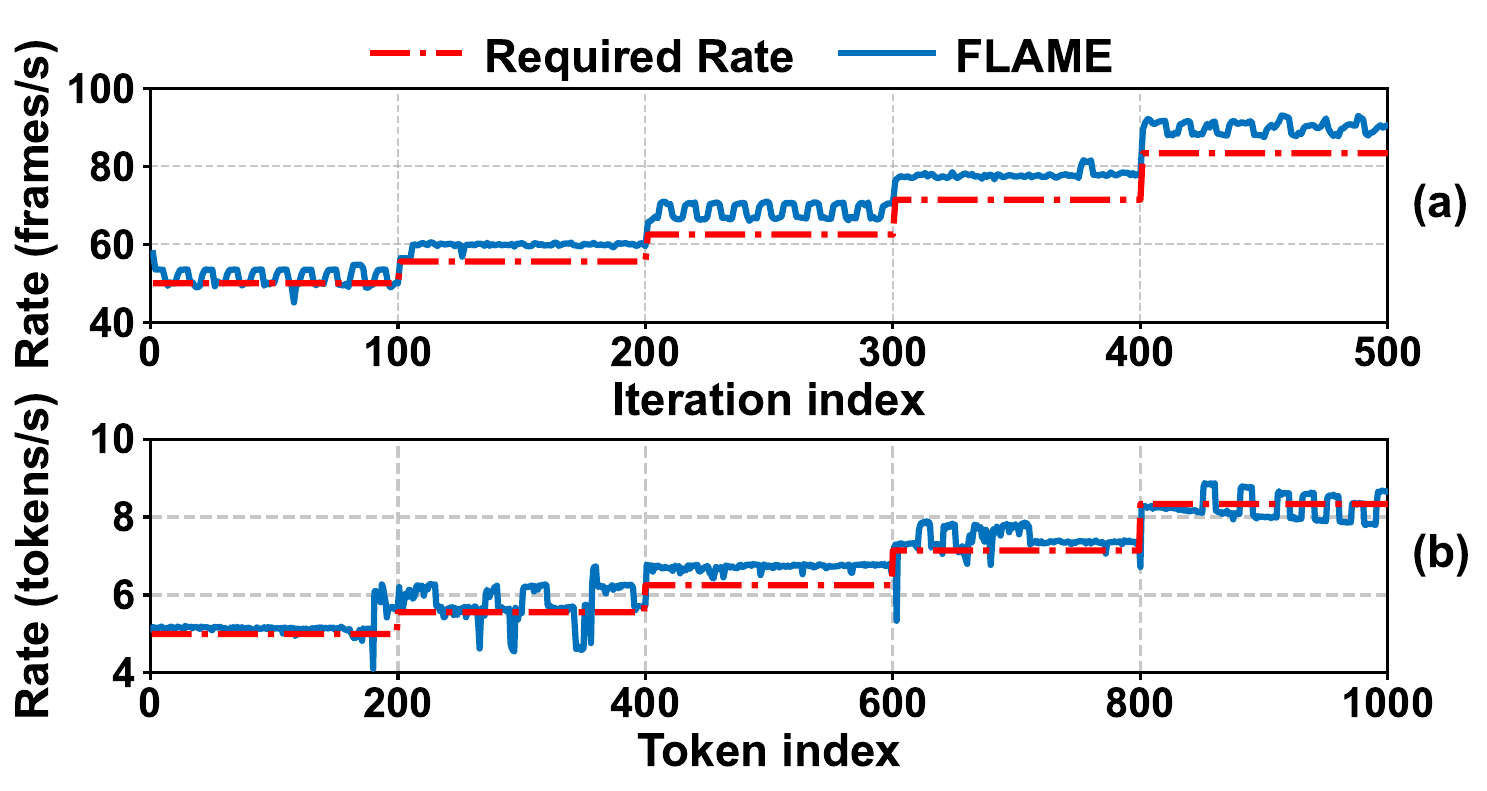}
    \end{center}
    \caption{Varying deadlines with (a) ResNet50 and (b) GPT2-large. The rate achieved (blue) higher than the required rate (red) means deadline requirement is met.}
    \label{fig:6_eval_micro_latency_trace}
    \vspace{-.1in}
\end{figure}

\textbf{Varying latency deadlines.} To test adaptability of \systemname-enhanced DVFS to various deadlines, we vary it in Figure~\ref{fig:6_eval_micro_latency_trace}. This governor itself does not require runtime deadline changes during its operation, and we change it for evaluation purposes only. For ResNet50, we tighten the deadline from 20 ms to 12 ms every 100 iterations (raising required frame rate from 50 to 83) in Figure~\ref{fig:6_eval_micro_latency_trace}(a). DVFS-\systemname responds swiftly, adjusting processors' frequencies to adapt to the deadline changes and achieving the rates slightly higher than the required rate to ensure not violating the deadline requirement. Similarly, for GPT2-large in Figure~\ref{fig:6_eval_micro_latency_trace}(b), shifting from 200 ms to 120 ms every 200 tokens (increasing required rate from 5.0 to 8.3 tokens/s) shows DVFS-\systemname adapting quickly as well. The latency curve for GPT2-large exhibits slightly larger fluctuations compared to ResNet50, primarily due to the more computationally intensive and variable \slm workloads. Overall, these results indicate DVFS-\systemname's responsive, precise control under various deadlines.

\begin{figure}[t]
    \begin{center}
        \includegraphics[width=0.97\linewidth]{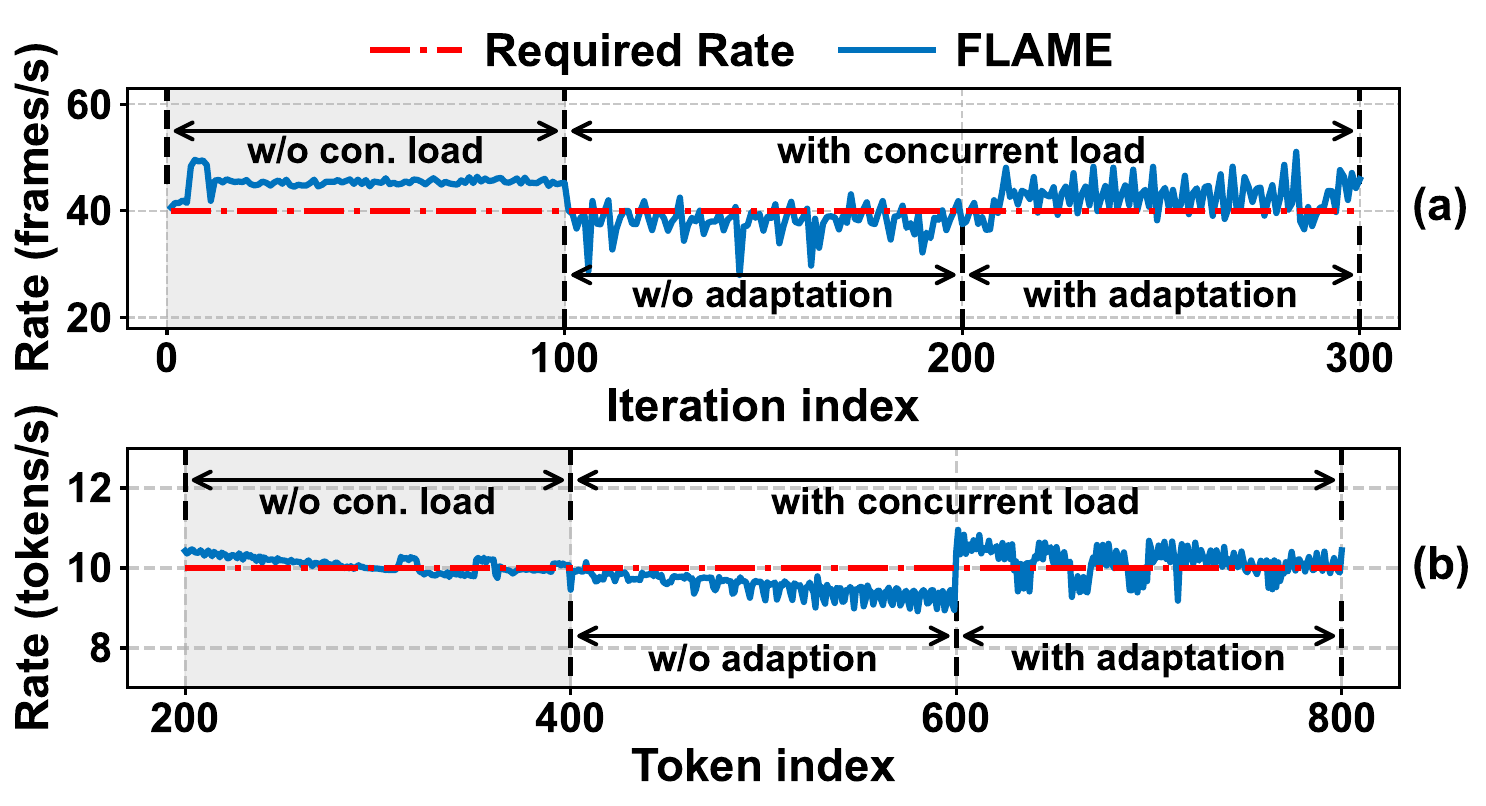}
    \end{center}
    \caption{Online adaptation for (a) ResNet50 and (b) GPT2-large with concurrent workloads.}
    \label{fig:6_eval_micro_concurrent}
    \vspace{-.1in}
\end{figure}

\textbf{Concurrent workloads.} To evaluate the online adaptation mechanism in \S\ref{sec:3_design_2} (to react to device workload dynamics), we ran MobileNet at 30 FPS as a concurrent task alongside ResNet50 and GPT2-large. For ResNet50 in Figure~\ref{fig:6_eval_micro_concurrent}(a), \systemname maintains the desired QoS (\ie the achieved rate higher than the required one) during iterations 0--100 without concurrent workload. When the background workload starts but the adaptation is disabled (iterations 100--200), a gradual degradation in the achievable rate is observed. Once the adaptation is activated (iterations 200--300), \systemname promptly compensates for the interference, restoring the rate no less than the desired level. A similar trend is observed for GPT2-large in Figure~\ref{fig:6_eval_micro_concurrent}(b). \systemname keeps the rate above the required rate without concurrent workload (tokens 200--400), experiences rate degradation when the additional workload appears (tokens 400--600), and effectively recovers once adaptation is enabled (tokens 600--800). These results show that online adaptation is effective in mitigating the impact of concurrent workloads and ensuring latency performance.

\begin{table}[b]
    \centering
    \caption{Profiling overhead of \systemname. The numbers in parentheses are the time needed for full profiling.}
    \label{tab:training_overhead}
    \resizebox{0.99\columnwidth}{!}{
    \begin{tabular}{lc|lc}
        \toprule
        \textbf{Model} & \textbf{Time} & \textbf{Model} & \textbf{Time} \\
        \midrule
        ResNet50       & 2 min (43 min)  & GPT2-large    &  2 min (113 h) \\
        VGG16          & 3 min (54 min)  & Qwen2-1.5B    & 2 min (151 h) \\
        DenseNet    & 6 min (102 min)  & Qwen2-7B      & 4 min (304 h) \\
        \bottomrule
    \end{tabular}}
\end{table}

\subsection{System Overhead}

We evaluate the system overhead on Jetson AGX Orin.

\textbf{Profiling overhead.} The accurate latency modeling of \systemname allows collecting only a very small subset of profiling data to train the estimator, \eg, the sampling interval is 4 and 90 for processor frequencies and context lengths as investigated in Figure~\ref{fig:6_eval_micro_diff_sample_intervals}. As summarized in Table~\ref{tab:training_overhead}, the profiling overhead of DNN models (400 iterations to smooth measurement jitter) and \slms (5 iterations) reduces from 43--102 minutes of full profiling to 2--6 minutes and from 113--304 hours to 2--4 minutes only, respectively.

\textbf{Inference latency and power.} Finally, we evaluate the runtime latency and power overhead of \systemname itself. Our measurements show that the latency of running \systemname for estimation takes only 3.32 ms, which is sufficient for real-time scheduling. In terms of power consumption, \systemname introduces an additional consumption of 0.05 W merely, orders of magnitude lower than the power consumption for executing DNNs or \slms. These results confirm that \systemname is a lightweight design.

\section{Related Work}
\label{sec:7_related}

\textbf{Latency estimation of model inference.} Accurate latency estimation is fundamental for deploying time-critical applications on the mobile edge, serving as an essential building block for model variant adaptation~\cite{ling2021rt}, neural architecture search~\cite{nnmeter}, and preemptive task scheduling~\cite{han2024pantheon}. While many existing designs rely on offline profiling, they typically assume static processor frequencies, an assumption that often fails in dynamic runtime environments. Although analytical and learning-based methods~\cite{dvfs_ICC, lyu2024predicting} attempt to mitigate this limitation, their coarse-grained modeling often results in significant estimation errors (\S\ref{sec:6_eval}). Other approaches, such as FLOPs-based designs~\cite{liu2018darts, real2019regularized}, attempt to scale latency estimations from small blocks to full models. More advanced frameworks like nn-Meter~\cite{nnmeter} incorporate graph optimizations to handle diverse architectures, while CoDL~\cite{codl} addresses heterogeneous scheduling between CPUs and GPUs. However, these methods are primarily tailored for mobile phones, a computing paradigm that is heavily CPU-centric due to mature tooling and general availability~\cite{chen2020deep}. In contrast, mobile edge (\eg, Jetson) relies on a distinct architecture where the GPU is the primary computational engine rather than a secondary accelerator alongside DSPs and NPUs~\cite{chen2020deep}. Consequently, phone-centric designs are ill-suited for the mobile edge. Furthermore, as existing methods primarily assume that operating frequencies of processors remain fixed as well, the \systemname design offers a novel perspective that could inspire future optimizations on mobile phone platforms.

While latency estimation is also performed on powerful platforms like cloud servers, the objectives differ from the mobile edge fundamentally. Cloud-based systems, utilizing GPU clusters, focus on predicting request arrival distributions~\cite{gujarati2020serving} to estimate batch processing latency~\cite{zhang2023shepherd}. This throughput-oriented paradigm is not applicable to the single-stream, latency-sensitive inference on mobile edge devices.

\textbf{DVFS on mobile edge.} Commercial mobile edge devices typically manage DVFS through efficient rule-based governors, such as \textit{schedutil} for CPUs and \textit{nvhost\_podgov} for GPUs~\cite{schedutil}. These governors rely on pre-defined heuristics to minimize power consumption~\cite{kim2018survey} but remain agnostic to inference latency. To bridge this gap, learning-based approaches have been proposed recently~\cite{gupta2019deep}, such as zTT~\cite{ztt} and GearDVFS~\cite{lin2023workload}. These approaches employ Reinforcement Learning (RL) to optimize Quality-of-Service (QoS) metrics, such as latency or processor utilization. In this paper, we employ \systemname to develop a new DVFS scheme to enable latency awareness, avoiding RL's slow convergence and outperforming RL-based solutions in both power efficiency and latency guarantees. A recent effort~\cite{dvfs_ICC} also attempts to make commercial DVFS latency-aware but its reliance on the standard inverse latency–frequency model leads to substantial errors under frequency scaling in real-world deployments.

\textbf{Time-critical mobile edge applications.} Mobile edge is increasingly pivotal in enabling next-generation technologies of autonomous vehicles~\cite{sun2024optimizing, shi2024soar}, embodied robotics~\cite{liu2024rl}, satellite computing~\cite{xu2022satellite}, UAVs~\cite{zhao2022dronesense}, etc. On-device model inference is crucial for a growing number of time-critical applications within these domains, from real-time object detection for collision avoidance in autonomous vehicles and UAVs~\cite{zhou2022integrated}, to efficient computing in outer space environment~\cite{9925218}, to latency-sensitive user interactions, including AI-enabled agents~\cite{wen2024autodroid} and embodied robotics powered by \slms~\cite{liu2025llm}. \systemname can  play an instrumental role in the development of these useful applications.

\section{Conclusion}
\label{sec:8_conclusion}

In this paper, we present \systemname, a framework that tames the complexity of asynchronous CPU-GPU coupling to enable precise, frequency-aware latency estimation of AI model inference on mobile edge devices. By shifting the paradigm from exhaustive brute-force profiling to sparse analytic modeling, \systemname eliminates the prohibitive overhead of characterizing model workloads, particularly emerging \slms. Our evaluation demonstrates that \systemname reduces profiling overhead by orders of magnitude while maintaining latency estimation error small. Furthermore, we leverage \systemname to enable a new latency-aware DVFS design, outperforming both commercial and state-of-the-art solutions in metrics of QoS guarantee and energy efficiency.

\section*{Acknowledgment}
This work is supported by the General Research Fund (GRF) grant from Research Grants Council of Hong Kong (CityU 11205624).

\let\oldbibliography\thebibliography
\renewcommand{\thebibliography}[1]{%
	\oldbibliography{#1}%
	\setlength{\itemsep}{1pt}%
}

\bibliographystyle{IEEEtran}
\bibliography{ref}

\end{document}